\documentclass[12pt,a4paper]{article}

\usepackage[margin=1.25in]{geometry}

\usepackage[natbibapa]{apacite}
\usepackage{graphicx}

\usepackage{amsmath}
\usepackage{amssymb}
\usepackage{breqn}
\usepackage{booktabs}
\usepackage{lineno}

\usepackage{verbatim}
\usepackage[T1]{fontenc}
\usepackage[font=small,labelfont=bf]{caption}
\usepackage[font=small]{subcaption}

\usepackage{authblk}

\title{Rate of Melt Ascent beneath Iceland from the Magmatic Response to 
Deglaciation}

\author[1]{Isarapong Eksinchol}
\author[1]{John F. Rudge}
\author[2]{John Maclennan}

\affil[1]{Bullard Laboratories, Madingley Road, Cambridge, CB3 0EZ}
\affil[2]{Department of Earth Sciences, University of Cambridge, Downing Street, Cambridge, CB2 3EQ}

\begin{document}
\maketitle



\section*{Key Points:}
\begin{itemize}
\item We model the magmatic response to the last deglaciation in Iceland with
finite melt ascent velocity.
\item The model results compared with observations suggest that the melt
ascent velocity is likely to be around $100$ $\mathrm{m/yr}$.
\end{itemize}

%
%


\begin{abstract}
Observations of the time lag between the last deglaciation and a surge in 
volcanic activity in Iceland constrain the average melt ascent velocity to 
be \mbox{$\geq50$ $\mathrm{m/yr}$}.
Although existing theoretical work has explained why the surge in eruption 
rates increased \mbox{$5$--$30$} fold from the steady-state rates 
during the last deglaciation, they cannot account for large variations 
of Rare Earth Element (REE) concentrations in the Icelandic lavas.
Lavas erupted during the last deglaciation are depleted in REEs by up 
to \mbox{$70\%$}; whereas, existing models, which assume 
instantaneous melt transport, can only produce at most \mbox{$20\%$} 
depletion.
Here, we develop a numerical model with finite melt ascent velocity and 
show that the variations of REEs are strongly dependent on 
the melt ascent velocity.
When the average melt ascent velocity is \mbox{$100$ $\mathrm{m/yr}$}, the 
variation of $\mathrm{La}$ calculated by our model is comparable to that 
of the observations.
In contrast, when the melt ascent velocity is \mbox{$1,000$ 
$\mathrm{m/yr}$} or above, the model variation of $\mathrm{La}$ becomes 
significantly lower than observed, which explains why previous models with 
instantaneous melt transport did not reproduce the large variations.
We provide the first model that takes account of the diachronous response
of volcanism to deglaciation.
We show by comparing our model calculations of the relative volumes of 
different eruption types (subglacial, finiglacial and postglacial) and the 
timing of the bursts in volcanic eruptions with the observations across 
different volcanic zones that the Icelandic average melt ascent velocity 
during the last deglaciation is likely to be \mbox{$\sim100$ $\mathrm{m/yr}$}.
\end{abstract}

%
%

%


%
%
%
%



\section{Introduction}

Iceland is located where a mantle plume meets the Mid-Atlantic Ridge 
\citep{White1992}.
The mantle plume and the spreading center
are responsible for the upwelling of the 
mantle underneath Iceland, which induces decompression melting in the 
upper mantle
\citep{McKenzie1988}.
This decompression melting produces magma that supplies the production of 
the Icelandic crust and volcanic eruptions.
Geological observations indicate that eruption rates in the volcanic zones 
of Iceland were significantly elevated during a burst of activity that 
took place after the end of the last major deglaciation 
\citep{Sigvaldason1992, Slater1998, Maclennan2002, Sinton2005, Maclennan2008, Eason2015}.
This period of high productivity, perhaps $30$ times or more, may have 
started about $15$ $\mathrm{kyrBP}$ and ended before $9$ $\mathrm{kyrBP}$.

The cause of the surge in eruption rates has been examined by 
\citet{Jull1996}.
In their model, post-glacial rebound induced by the last deglaciation 
increases the rate of pressure drop in the upper mantle by up to $50$-fold 
from the steady-state value.
This increase in the decompression rate significantly increases the 
melting rate in the upper mantle, which leads to greater melt supply.

\citet{Jull1996} also showed that the decompression rate due to 
post-glacial rebound has its maximum value at the surface and decays 
exponentially with depth.
This means that the additional melt production in the mantle during the 
deglaciation occurs mostly at shallow depths.
The melts generated at these depths are depleted in Rare Earth Elements 
(REEs).
These additional melts produced during the deglaciation will therefore 
dilute the concentrations of REEs in the aggregated melts.
By assuming that the melt transport is instantaneous, \citet{Jull1996} 
calculated that
the REE concentrations in the melts decrease by around $20\%$ during the 
last deglaciation compared to melts generated at other times when the 
ice-load is thought  to have been close to steady-state.
However, geological observations indicate that lavas erupted during the 
surge in volcanic eruption rates are depleted in REEs by up to 
$\approx70\%$
\citep{Slater1998, Maclennan2002, Sinton2005, Maclennan2008, Eason2015},
which is significantly higher than that calculated by 
\citet{Jull1996}.

\citet{Slater1998} attempted to account for this mismatch by developing an 
inverse model similar to that of \citet{McKenzie1991}, which used the 
observed variations of REE concentrations to constrain the melt 
productivity function.
\citet{Slater1998} showed that there exists a melt productivity function 
that matches the \citet{Jull1996} theoretical variations of the REE 
concentrations with the geological data.
However, in order for such a melt productivity function to exist, model 
parameters had to be modified including reducing the initial ice sheet 
radius to $90$ $\mathrm{km}$, which is significantly smaller than the 
likely radius of the ice sheet as inferred from observational studies 
\citep{Sigmundsson1991, Hubbard2006, Licciardi2007, Petursson2015, Patton2017}.

\citet{Maclennan2002} previously used the relative timing of the last 
deglaciation and the surge in volcanic eruption rates in the Northern 
Volcanic Zone (NVZ) of Iceland to estimate that the melt ascent velocity 
is at least $50$ $\mathrm{m/yr}$.
The Icelandic melt ascent velocity during the mid-Holocene has also been 
estimated from a time lag of $\approx 600$ $\mathrm{yr}$ 
\citep{Swindles2017} obtained from the cross-correlation between the 
Icelandic volcanic eruption rates and the change in the atmospheric 
circulation pattern indicated by sodium concentrations in Greenland Ice 
Sheet Project 2 (GISP2).
This $\approx 600$ $\mathrm{yr}$ time lag gives an estimated Icelandic 
melt ascent velocity of $\approx 50$--$100$ $\mathrm{m/yr}$ during the 
mid-Holocene.

Here, we investigate how the melt ascent velocity in the mantle and the 
crust influences the variations of REE concentrations.
By incorporating a finite rate melt transport model into the model of 
\citet{Jull1996}, we show that variations of REE concentrations depend 
significantly on the melt ascent velocity.
With an appropriate melt ascent velocity, our model demonstrates that the 
model variations of REE concentrations can be matched with those observed 
geologically.
While \citet{Jull1996} used a very simple ice-load model with a constant 
ice radius, our model combines an ice-load history with melt generation 
and transport and therefore enables prediction of the volumes and 
compositions of melt that are erupted either subglacially or in ice-free 
settings.
This feature of the model allows for direct comparison with geological 
observations, which use edifice geomorphology and volcanic facies analysis 
to determine whether an eruption is subglacial or not.
Therefore, not only does this work help us understand how melt transport 
affects REE concentrations and eruption types in different places on 
Iceland, but also
it can
be useful as a tool to constrain the melt 
transport rate.

In the next section, we will begin by considering how the mantle flow 
responds to deglaciation and use this response to calculate the melting 
rate and the compositions of the melts generated.
We then transport the melts produced at depth to the surface to calculate 
the eruption rate together with the composition of the erupted lavas.
Finally, we compare our numerical results to the observational data in 
order to constrain the melt ascent velocity.

\newpage
\section{Model}

We follow the modelling of \citet{Jull1996} with the following key 
differences:
\begin{enumerate}
\item[1.]
While \citet{Jull1996} used an ice-load with a constant radius,
our ice sheet behaves like a gravity current with time-dependent radius 
and thickness.
\item[2.]
Our ice-load input consists of multiple deglaciation stages beginning at 
$23.0$ to $10.5$ $\mathrm{kyrBP}$ designed to capture key features of 
ice-sheet reconstructions.
\item[3.]
We neglect the elastic response of the solid mantle.
\item[4.]
We assume finite melt ascent velocity.

\end{enumerate}
Numerical parameters used as inputs into the model are listed in Table 
\ref{tab:parameters}.

\begin{table}[ht]
	\centering
		\caption{Parameter values for calculations.}
		\label{tab:parameters}
		\begin{tabular}{r l r l}
			\toprule
			\textbf{Parameter}	&\textbf{Meaning}	
					&\textbf{Value}	
&\textbf{Dimensions}\\
			\midrule
			$D^{\mathrm{La}}$	& $\mathrm{La}$ partition coefficient	& $0.010$				& $1$\\			
			$-\left(\partial F/\partial P\right)_S$	& isentropic melt productivity		& 10	& $\mathrm{wt\%/GPa}$\\
			$g$					& gravitational acceleration			& $9.82$				& $\mathrm{m/s^2}$\\
			$P_{sol}$			& solidus pressure						& $3.5$					& $\mathrm{GPa}$\\
			$U_0$				& plate half-spreading rate				& $10$					& $\mathrm{mm/yr}$\\
			$z_c$				& crustal thickness						& $20$					& $\mathrm{km}$\\
			$\alpha$			& ridge angle							& $45$					& $\mathrm{deg}$\\
			$\eta$				& mantle viscosity						& $8.0\times 10^{18}$	& $\mathrm{Pa\,\,s}$\\
			$\rho_i$			& density of ice						& $900$					& $\mathrm{kg/m^3}$\\
			$\rho_l$			& density of melt						& $2900$				& $\mathrm{kg/m^3}$\\
			$\rho_s$			& density of solid mantle				& $3300$				& $\mathrm{kg/m^3}$\\
			$\tau_B$			& yield stress of ice					& $100$					& $\mathrm{kPa}$\\[0.5ex]\bottomrule
		\end{tabular}
	
\end{table}

Numerical values of the plate half-spreading rate ($U_0$), crustal thickness
($z_c$), ridge angle ($\alpha$) and mantle viscosity ($\eta$) are the same as
in \protect\citet{Jull1996}.
$z_c=20$ $\mathrm{km}$ is at the lower bound of the
\protect\citet{Darbyshire2000}
estimates ($20$--$37$ $\mathrm{km}$) because our study areas are relatively far
($>100$ $\mathrm{km}$) from the mantle plume center.
Numerical values of $U_0=10$ $\mathrm{mm/yr}$ and
$\eta=8.0\times 10^{18}$ $\mathrm{Pa\,\,s}$
are similar to the \protect\citet{Arnadottir2009} estimates.
The density of ice ($\rho_i$) is in the range of
$830$--$917$ $\mathrm{kg/m^3}$ in \citet{Paterson1994}.
The densities of melt ($\rho_l$) and of solid mantle ($\rho_s$) follow
\citet{Katz2003}.
Sources of the remaining numerical parameters will be mentioned later.

\newpage
\subsection{Glacial Load} \label{sec:glacialLoadInput}
Due to limited geological records of the ice sheet, it is not 
straight-forward to reconstruct the details of the shape of the ice-sheet 
during the last deglaciation 
\citep{Hubbard2006, Patton2017}.
In \citet{Jull1996}, the ice sheet was assumed to have axisymmetric 
parabolic shape with a constant radius of $180$ $\mathrm{km}$.
Here, we modify the ice sheet to be an axisymmetric viscous gravity 
current \citep{Paterson1994} with glacier terminus retreating during the 
deglaciation. This is a more reasonable representation of the actual ice 
sheet, allowing the spatial variations of the volcanic response to be 
examined more accurately.

In an axisymmetric gravity current ice model
\citep{Huppert1982, Paterson1994},
the differential thickness of ice along the radial direction induces a
radially-inward basal shear stress.
At the yield strength limit of ice,
the basal shear stress is uniform and is equal to the yield stress of ice.
When the stress exceeds the yield strength limit,
ice deformation and sliding will occur.
These processes will re-adjust the ice sheet shape until it returns back 
to within its yield strength limit.
We assume here that the time scale of the ice deformation and sliding
(when the stress exceeds the yield strength) is short compared to the
time scale of the deglaciation.
That is, the glacier is assumed to always be within its yield strength 
limit and the thickness of ice $h(r,t)$ as a function of radial distance 
$r$ and time $t$ follows
\citep{Huppert1982, Paterson1994}
\begin{linenomath*}
\begin{align}
	h(r,t) = 
	\begin{cases}
	h_m(t) \sqrt{1 - \frac{r}{r_m(t)}}, \, & 0 \leq r \leq r_m(t),\\
	0,              & \text{otherwise,}
	\end{cases}
	\label{eq:iceProfile}
\end{align}
\end{linenomath*}
where
\begin{linenomath*}
\begin{align*}
	h_m(t)
	& = \left(\frac{15}{8\pi} V(t)\right)^{\frac{1}{5}} \, 
\left(\frac{2\,\tau_{B}}{\rho_{i} \, g}\right)^{\frac{2}{5}} \, ,
	\\
	r_m(t)
	& = \left(\frac{15}{8\pi} V(t)\right)^{\frac{2}{5}} \, 
\left(\frac{\rho_{i} \, g}{2\,\tau_{B}}\right)^{\frac{1}{5}} \, .
\end{align*}
\end{linenomath*}
$h_{m}(t)$ is the thickness of ice at the center, 
$r_{m}(t)$ is the radial extend of ice, $V(t)$ is the volume of ice, 
$\rho_{i}$ is the density of ice and $\tau_{B}$ is the yield stress of ice.

The numerical value of $\tau_{B}= 100$ $\mathrm{kPa}$ we use here 
\citep{Paterson1994} gives ice sheet dimensions that closely resemble that 
of the Late Weichselian Icelandic ice sheet
\citep{Sigmundsson1991, Hubbard2006, Licciardi2007, Petursson2015, Patton2017}
and can reproduce the ice radius of $180$ $\mathrm{km}$ 
together with $2$ $\mathrm{km}$ ice thickness used previously in 
\citet{Jull1996}.

The time evolution of the ice coverage as an input into our model follows 
approximately that of \citet{Patton2017}.
We set the input deglaciation to consist of three stages during which the 
ice volume decreases linearly with time and the ice volume stays constant 
during two intermissions between these three deglaciation stages.
Time $t=0$ in the model corresponds to the present (AD 1950).

\begin{figure}[ht!]
	\begin{center}
		\includegraphics[width=1.0\textwidth]{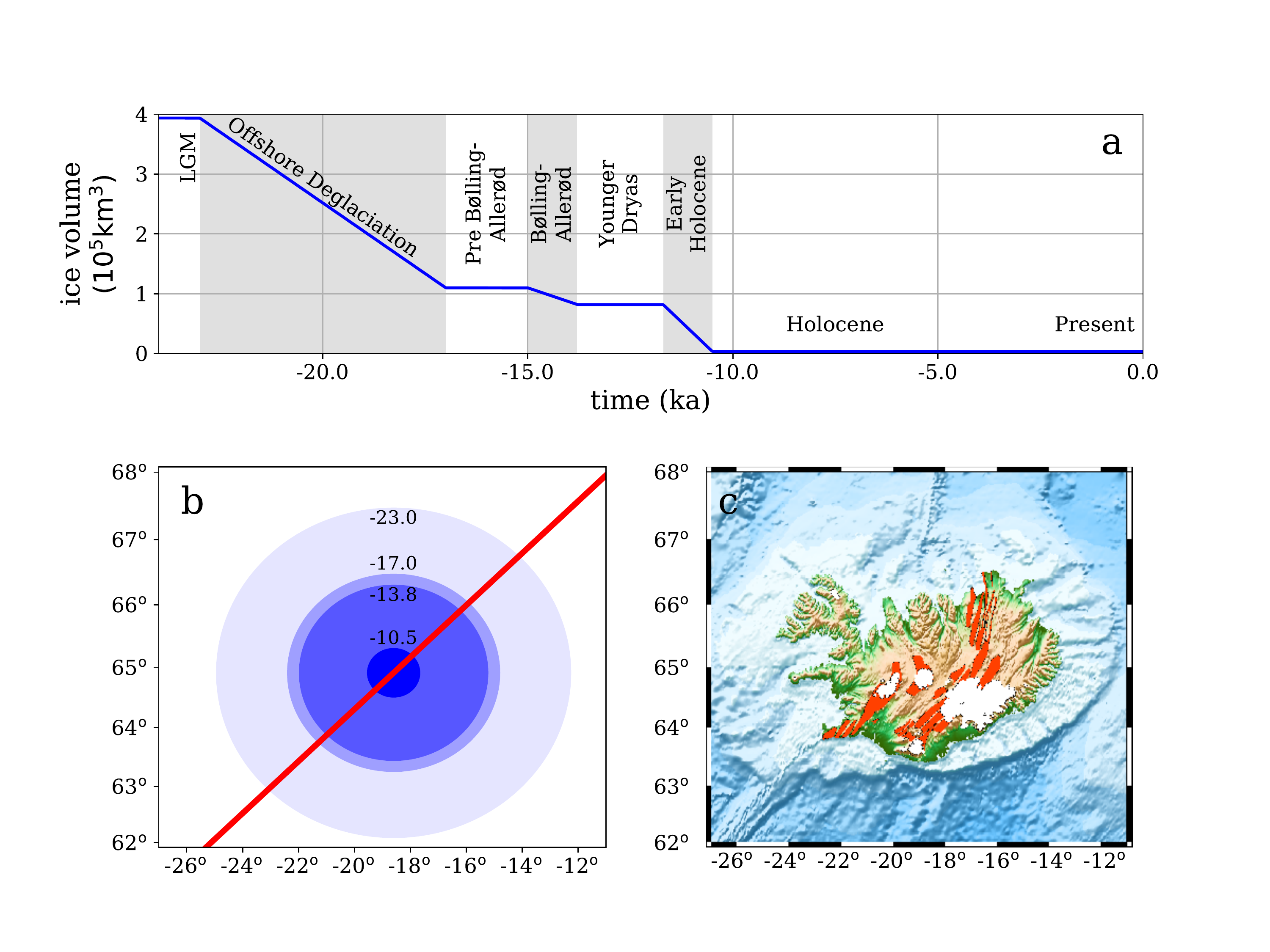}
\caption{(a) The modelling-input ice volume history with grey bars 
labelling deglaciation periods.
(b) Top-view of the model rift (red line) with snapshots of the 
modelling-input ice coverage (blue circles) at time $t=$ $-23.0$, $-17.0$, 
$-13.8$ and $-10.5$ $\mathrm{kyr}$ as labelled at the edges of the 
circles. (c) Map of Iceland with the same length scale as in (b) showing 
locations of the fissure swarms (red color) where the plate spreading takes 
place and the current glaciers in white color.
See Supporting Information for animation of the model ice coverage map.}
		\label{fig:iceMap}
	\end{center}
\end{figure}

We set the initial ice load to have a radius of $300$ $\mathrm{km}$ 
covering the whole of Iceland and most of the continental shelf.
The ice volume is held constant until $t=-23.0$ $\mathrm{kyr}$ when the 
first stage of deglaciation (we refer to as the Offshore Deglaciation) 
begins.
The Offshore Deglaciation terminates at the shoreline with ice radius of 
$180$ $\mathrm{km}$ at time $t=-17.0$ $\mathrm{kyr}$ followed by a pause 
of $2.0$ $\mathrm{kyr}$.
Next, the second stage of deglaciation (B{\o}lling-Aller{\o}d) proceeds 
from time $t=$ $-15.0$ to $-13.8$ $\mathrm{kyr}$ during which the ice 
radius decreases from $180$ $\mathrm{km}$ to $160$ $\mathrm{km}$.
Then, the deglaciation pauses for $2.1$ $\mathrm{kyr}$, corresponding 
approximately to the Younger-Dryas.
The final stage of deglaciation (Early Holocene) takes place between time 
$t=$ $-11.7$ and $-10.5$ $\mathrm{kyr}$ with the ice sheet retreating from 
radius of $160$ $\mathrm{km}$ to $45$ $\mathrm{km}$, which is 
approximately the current size of the Vatnaj\"okull ice sheet.
The timeline of the modelling-input deglaciation is comparable to the last 
deglaciation in Iceland
\citep{Sigmundsson1991, Maclennan2002, Hubbard2006, Licciardi2007, Petursson2015, Patton2017}
and is summarised in Figure \ref{fig:iceMap}, Table 
\ref{tab:iceTimelines} and the Supporting Information.

\begin{table}[ht]
\centering
\caption{Timeline of the model ice sheet.}
\label{tab:iceTimelines}
\begin{tabular}{r c l l}
	\toprule
	\textbf{Geological Period}	&\textbf{Time (kyrBP)}		&\textbf{Ice Radius (km)}		&\textbf{Average Ice Thickness (km)}\\
	\midrule
	Last Glacial Maximum		& before $23.0$				& $300.0$ constant				& $1.39$ constant\\
	Offshore Deglaciation		& $23.0 \rightarrow 17.0$	& $300.0 \rightarrow 180.0$		& $1.39 \rightarrow 1.08$\\
	Pre B{\o}lling-Aller{\o}d	& $17.0 \rightarrow 15.0$	& $180.0$ constant				& $1.08$ constant\\
	B{\o}lling-Aller{\o}d		& $15.0 \rightarrow 13.8$	& $180.0 \rightarrow 160.0$		& $1.08 \rightarrow 1.02$\\
	Younger Dryas				& $13.8 \rightarrow 11.7$	& $160.0$ constant				& $1.02$ constant\\
	Early Holocene				& $11.7 \rightarrow 10.5$	& $160.0 \rightarrow 45.0$		& $1.02 \rightarrow 0.54$\\
	Holocene					& $10.5 \rightarrow$ now	& $\,\,\,45.0$ constant			& $0.54$ constant\\					
	\bottomrule
\end{tabular}
\end{table}

\subsection{Mantle Flow}
Similar to the assumption made by \citet{Jull1996},
in steady state, the spreading ridge induces passive upwelling of the 
mantle, which we assume to follow corner flow
\citep{Batchelor2000, Spiegelman1987}.
Active upwelling induced by the mantle plume can also increase the
melt production rate.
However, the geological data in our study come from regions that are at
least $\sim100$ $\mathrm{km}$ away from the plume center.
We therefore assume that the active upwelling is insignificant here.

\begin{figure}[ht!]
	\begin{center}
		\includegraphics[width=0.75\textwidth]{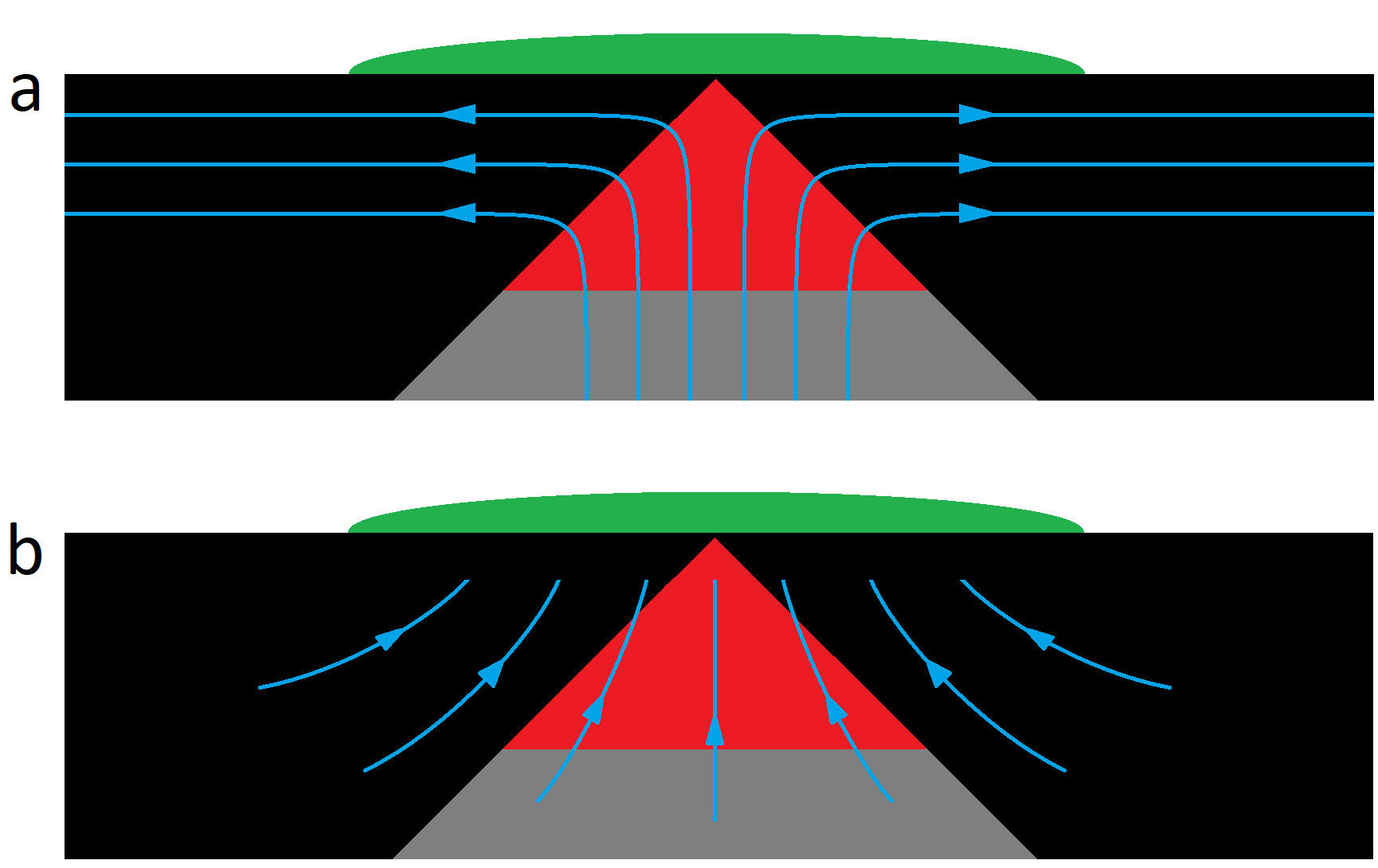}
\caption{Simplified diagrams illustrating the
solid mantle
streamlines of (a) corner flow and (b) Glacial Isostatic Adjustment (GIA)
in a vertical plane passing through the center of ice
perpendicular to the ridge axis.
In steady state, the decompression melting comes from the upwelling of the 
mantle due to the spreading ridge (with half-spreading rate of $10$
$\mathrm{mm/yr}$)
and the mantle plume (which we assume to be insignificant
in the studied areas).
During deglaciation, the GIA further increases the mantle upwelling
rate and hence the decompression melting rate.}
		\label{fig:mantleFlow}
	\end{center}
\end{figure}

The glacial load on the surface affects the pressure in the mantle 
underneath.
During deglaciation, the surface load drops, which leads to
an increased
mantle decompression melting rate from that induced 
by the steady state passive corner flow (Figure \ref{fig:mantleFlow}).

To calculate the effect of deglaciation on mantle melting rate,
we first note that the Maxwell relaxation time ($\tau_M = \eta/\mu \approx 
10^1$ $\mathrm{yrs}$) is much shorter than the viscous characteristic time 
($\tau_v = 2\eta k/\rho_s g \approx 10^3$ $\mathrm{yrs}$) where
$\mu$ is the elastic modulus, $k$ is wave number and other variables are 
as defined in Table \ref{tab:parameters}. 
This means that the elastic deformation in the  viscoelastic mantle model 
used in \citet{Jull1996}
is negligible and the deformation in the mantle is dominated by the 
viscous response.
We therefore model the mantle as a viscous half-space incompressible fluid 
and the elastic thickness of the Icelandic lithosphere is assumed to be 
negligible.
When using the same modelling inputs as in \citet{Jull1996},
our numerical model yields the same results as those in \citet{Jull1996}, 
which verifies our assumption that the elastic deformation is 
insignificant.

The boundary conditions on the surface of the half-space mantle are that 
the normal stress is equal to the pressure from the weight of the ice load 
and that the shear stress is negligible compared to the normal stress.
We obtain semi-analytical solutions to the mantle flow in cylindrical 
coordinates in response to the glacial load as provided in Appendix
\ref{appendix:mantleFlow} together with the corner flow solutions.

\newpage
\subsection{Decompression Melting in the Mantle}
Mantle upwelling is sufficiently fast that the heat loss due to conduction 
is negligible. Therefore, the decompression melting in the mantle is 
adiabatic and the mantle melting rate $DF/Dt$ can be calculated by 
\citep{Jull1996}
\begin{linenomath*}
\begin{equation}
	\frac{DF}{Dt} = \left(
	\frac{\partial F}{\partial P}
	\right)_S
	\, \frac{DP}{Dt}
	\label{eq:melting&decompressionRate}
\end{equation}
\end{linenomath*}
where
$F$ is the degree of melting by
mass
fraction
relative to the initial mass of the solid mantle,
$\left(\partial F/\partial P\right)_S$ is the isentropic melt productivity 
of the mantle
and $\frac{D}{Dt}$ is the convective derivative following
the solid mantle trajectories.

The rate of mass production of melt per unit volume as a function of space
and time is assumed to follow
\protect\begin{linenomath*}
\begin{align}
	\Gamma(\boldsymbol{x},t)
	&= \rho_s \frac{DF}{Dt} 
	\nonumber \\
	&= \rho_s\left(\frac{\partial F}{\partial P}\right)_S \,\frac{DP}{Dt}.
	\label{eq:meltProductionRateField}
\end{align}
\end{linenomath*}

The isentropic melt productivity depends on several factors including the 
composition of the mantle, temperature and pressure \citep{McKenzie1984}. 
In numerical calculations, using different melt productivity functions 
will result in different profiles of depth-dependent mantle melting rate, 
and different eruptive REE concentrations \citep{Slater1998}. To 
investigate the effect of magma transport solely without the effect of 
melt productivity function on the eruptive REE concentrations, we use a 
constant isentropic melt productivity (Table \ref{tab:parameters}) and the 
degree of melting as a function of pressure follows a simple linear 
relation
\begin{linenomath*}
\begin{equation}
	F(P) = -\left(
	\frac{\partial F}{\partial P}
	\right)_S
	\, \left( P_{sol} - P \right).
	\label{eq:degMelting}
\end{equation}
\end{linenomath*}
Our choice of solidus pressure and melt productivity (in Table 
\ref{tab:parameters}) gives a melt productivity function that closely 
resembles that obtained from the melt parametrisation of \citet{Katz2003} 
at $1500$ $\mathrm{^oC}$ potential temperature.

\protect\citet{Sims2013} have shown that the temporal variability of
isotope ratios in lavas erupted during the last deglaciation in northern
Iceland provide evidence for a lithologically heterogeneous mantle source
beneath Iceland.
We investigate the effect of mantle heterogeneities by comparing our
simple homogeneous mantle model results to that of the pMELTS modelling
\protect\citep{Ghiorso2002, Smith2005} of a bi-lithological mantle
as used in \protect\citet{Rudge2013}.
We show these results in Supporting Information that both mantle models
yield the same conclusions for the rate of melt ascent.
Our model is not very sensitive to the mantle heterogeneities because
the model calculations do not involve isotopic composition.

Melts generated in the mantle have to be transported to the surface before 
they erupt.
We assume that the effects of finite melt transport rate can be approximated
by sampling the melt production rate field
(equation \protect\eqref{eq:meltProductionRateField})
with a time-lagged sampler.
To the leading order, we assume that the vertical component of the melt
velocity is constant $=v_t$.
In this case, the time taken for melt produced at location $(x,y,z)$ in the
mantle to ascend to the surface is $\Delta t=|z|/v_t$, where $|z|=-z$
($\because z<0$ below the Earth's surface).
That is, melt that reaches the surface at time $t$ is assumed to have been
produced at time $t'=t-|z|/v_t$ in the past.
Therefore, the total mass flux of melt supply to the crust at time $t$ is
\protect\begin{linenomath*}
\begin{align}
	\dot{M}(t) &= \int_{\mathcal{V}}
		\Gamma\left(\boldsymbol{x},t-\frac{|z|}{v_t}\right) \,dV,
	\label{eq:massFluxOfMelt}
\end{align}
\end{linenomath*}
which is the integral of all the instantaneous melts produced in the melting
region $\mathcal{V}$; however, the melts added from depth $|z|$ are assumed to
have been produced at time $t'=t-|z|/v_t$ in the past.

The total volume flux $\dot{V}$ of melt supply to the crust at time $t$ can be
calculated from the mass flux:
\protect\begin{linenomath*}
\begin{align}
	\dot{V}(t) &= \frac{\dot{M}(t)}{\rho_l}
	\nonumber \\	
	&= \frac{1}{\rho_l}\int_{\mathcal{V}}
		\Gamma\left(\boldsymbol{x},t-\frac{|z|}{v_t}\right) \,dV.
	\label{eq:volumeFluxOfMelt}
\end{align}
\end{linenomath*}

\newpage
\subsection{REE Concentrations}
We simplify the model by assuming that the concentration
$c^i_l$
of a highly incompatible element
$i$
with partition coefficient $D^i$ in the instantaneous melt can be calculated
based on modal fractional melting \citep{Shaw1970}
\protect\begin{linenomath*}
\begin{equation}
	\frac{c^i_l}{c^i_{s0}} = \frac{1}{D^i}\left( 1 - F \right)^{\frac{1}{D^i} - 1}
	\label{eq:instantaneousConcentration}
\end{equation}
\end{linenomath*}
where
$c^i_{s0}$
is the concentration of the element in the initial source.

Equation \eqref{eq:instantaneousConcentration} gives the instantaneous 
concentration as a function of the degree of melting
$c^i_l = c^i_l(F)$.
The degree of melting as a function of pressure $F = F(P)$ is known from 
equation \eqref{eq:degMelting} and the pressure as a function of position 
and time
$P = P(\boldsymbol{x},t)$
is known from equation \eqref{eq:rebound}. We can 
therefore combine these equations to calculate at any location in the 
mantle at any time the instantaneous concentration
$c^i_l = c^i_l(\boldsymbol{x},t)$
in the melt generated.
The bulk partition coefficient of $\mathrm{La}$ (Table 
\ref{tab:parameters}) is assumed to follow that in \citet{Workman2005}.

This very simplified melting modelling of $\mathrm{La}$ gives results that 
are not significantly different from those (shown in Supporting Information)
obtained from a more elaborate model of mantle melting used in
\citet{Rudge2013} because highly incompatible elements (such as $\mathrm{La}$)
partition into melts almost completely near the solidus intersection in the
garnet field.

Given the concentration (by mass) $c^i_l$
of a trace element $i$ in the instantaneous melt as a function of space and time,
the total mass flux of the trace element $i$ in the melt supply to the crust is
\protect\begin{linenomath*}
\begin{align}
	\dot{M_i}(t) &= \int_{\mathcal{V}}
		c^i_l \, \Gamma\left(\boldsymbol{x},t-\frac{|z|}{v_t}\right) \,dV
	\label{eq:massFluxOfElement}
\end{align}
\end{linenomath*}
where $c^i_l$ is calculated at point $(\boldsymbol{x},t-\frac{|z|}{v_t})$.

Similar to the volume flux of the whole melt defined in equation
\protect\eqref{eq:volumeFluxOfMelt},
we define
the total ``volume'' flux of a trace element $i$ in the melt supply to the crust
as
\protect\begin{linenomath*}
\begin{align}
	\dot{V_i}(t) &= \frac{\dot{M_i}(t)}{\rho_l}
	\nonumber \\
		&= \frac{1}{\rho_l} \int_{\mathcal{V}}
		c^i_l \, \Gamma\left(\boldsymbol{x},t-\frac{|z|}{v_t}\right) \,dV.
	\label{eq:volumeFluxOfElement}
\end{align}
\end{linenomath*}
Following these definitions, the mean concentration of the element $i$
in the melt supply to the crust at time $t$ is
\protect\begin{linenomath*}
\begin{align}
	\bar{c}^i_l(t) = \frac{\dot{M}_i(t)}{\dot{M}(t)}
	= \frac{\dot{V}_i(t)}{\dot{V}(t)}.
	\label{eq:concentrationFluxOfElement}
\end{align}
\end{linenomath*}

\newpage
\section{Results and Discussion}

\subsection{Decompression Melting and Eruption Rates}

The numerical methods we use for the calculations are discussed in 
Appendix \ref{appendix:numericalMethods}.
Figure \ref{fig:contourDpDt} illustrates snapshots of the decompression 
rate in the mantle from the model when the ice load history follows the 
timeline given in Section \ref{sec:glacialLoadInput}.

\begin{figure}[ht!]
	\begin{center}
		
\includegraphics[width=1.00\textwidth]{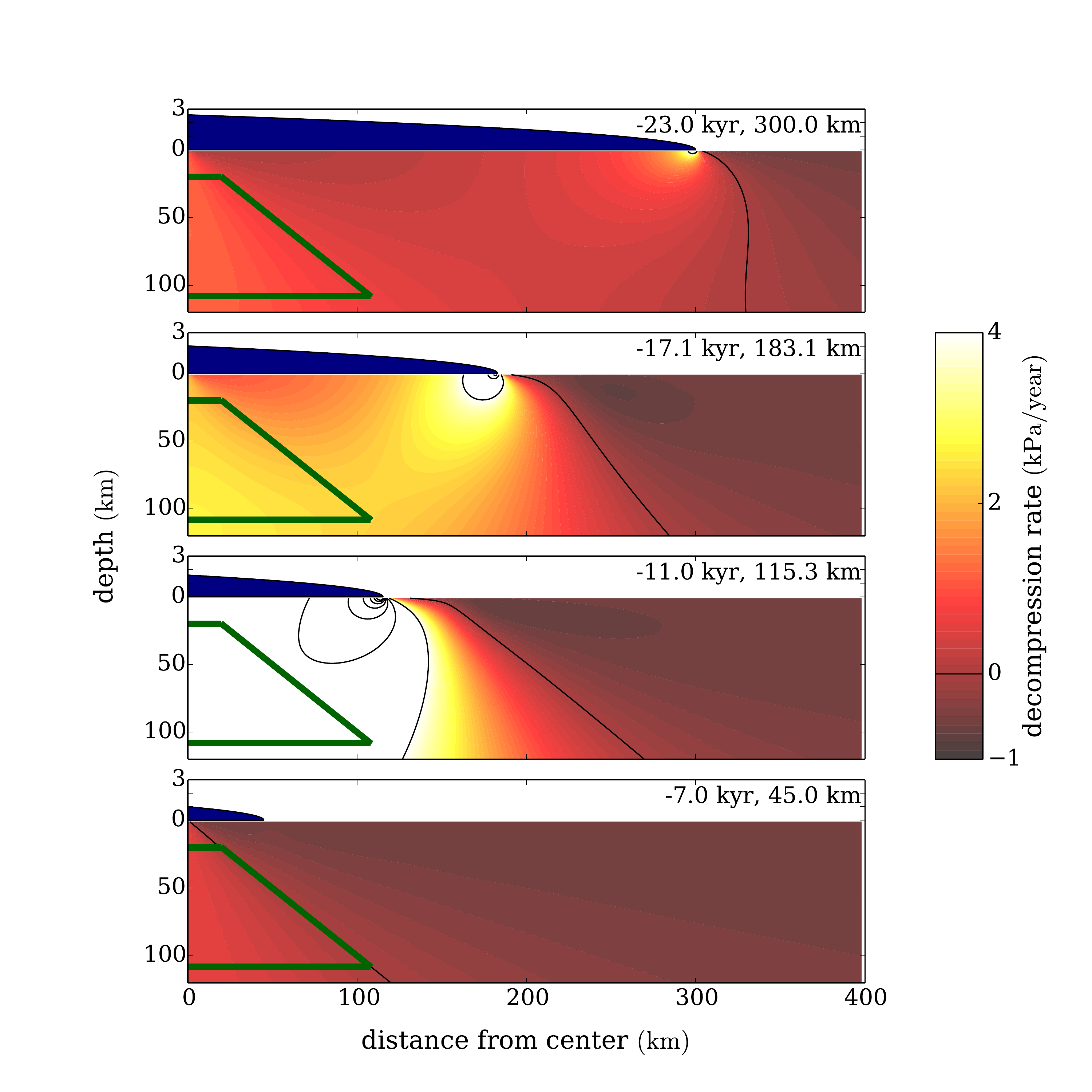}
\caption{Snapshots of the decompression rate in a vertical plane passing 
through the center of ice perpendicular to the ridge axis
(ridge axis as red line in Figure \protect\ref{fig:iceMap}b)
induced by the 
post-glacial rebound and the corner flow.
Black contour lines are separated at equal intervals of $4$ 
$\mathrm{kPa/yr}$.
The time and ice radius are shown in the upper right corner of each panel.
The deglaciation is assumed to take place between time $t=23.0$--$10.5$ 
$\mathrm{kyrBP}$ with two pauses in between at $t=17.0$--$15.0$ 
$\mathrm{kyrBP}$ and $13.8$--$11.7$ $\mathrm{kyrBP}$ during which the ice 
volume stays constant (see Section \ref{sec:glacialLoadInput} for details).
The ice load profile (navy blue color) is drawn on top of the mantle with 
$15\times$ vertical exaggeration.
Boundaries of the mantle melting region are outlined by the dark green 
lines.
Animation of the decompression rates in the mantle is provided in the 
Supporting Information.}
		\label{fig:contourDpDt}
	\end{center}
\end{figure}

\begin{figure}[ht!]
	\begin{center}
		
\includegraphics[width=1.00\textwidth]{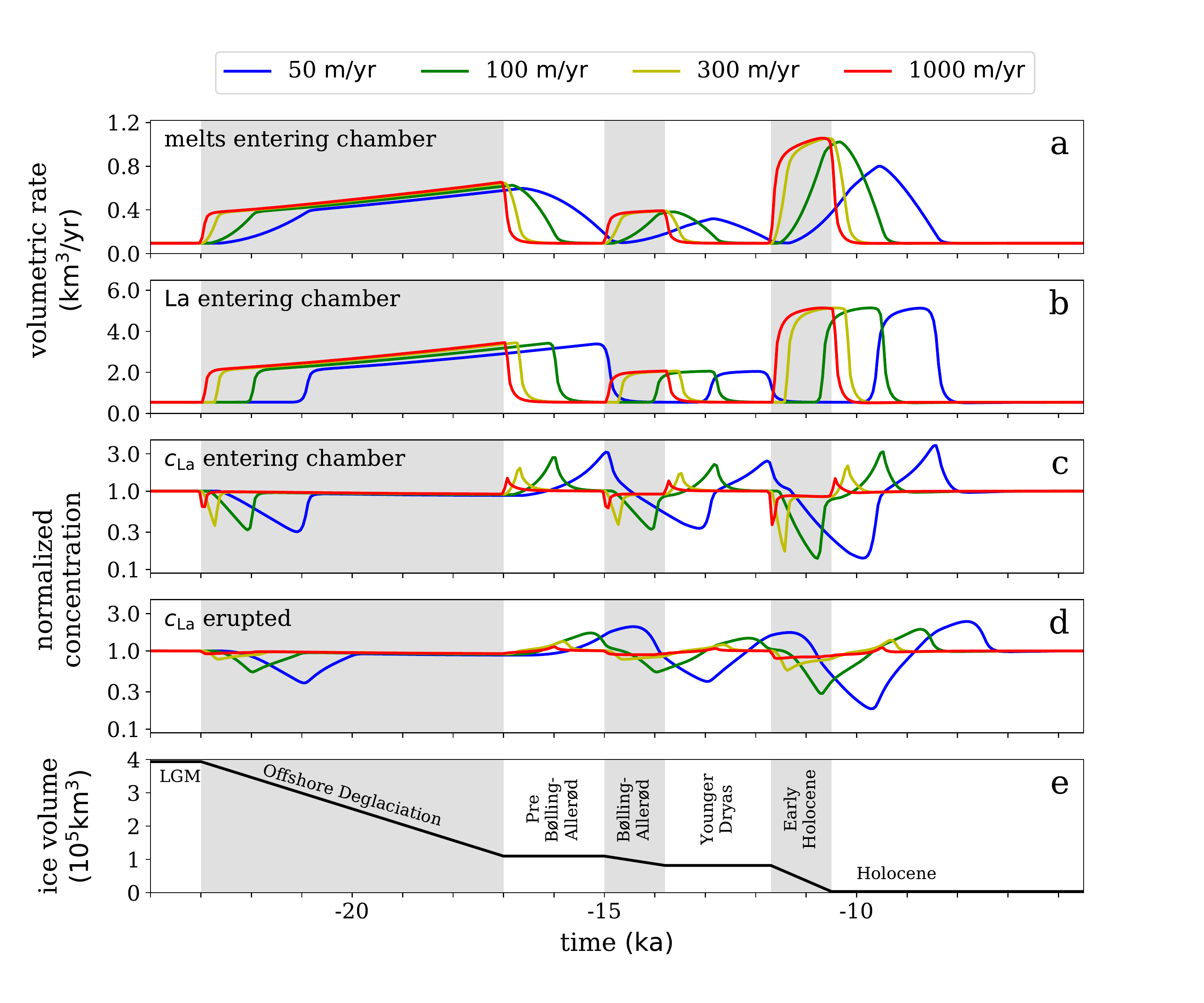}
\caption{a) volumetric rate of melt supply to the crustal chamber
(equation \protect\eqref{eq:volumeFluxOfMelt}).
b) 
volumetric rate of $\mathrm{La}$ supply to the crustal chamber
(equation \protect\eqref{eq:volumeFluxOfElement})
normalized to the $\mathrm{La}$ concentration in the source.
c) concentration of $\mathrm{La}$ in melt supply to the crustal chamber
(equation \protect\eqref{eq:concentrationFluxOfElement})
normalized to the steady-state $\mathrm{La}$ concentration.
d) concentration of $\mathrm{La}$ in erupted lavas normalized to the 
steady-state $\mathrm{La}$ concentration.
e) modelling-input ice load volume.
(c) is the ratio of the $\mathrm{La}$ volume (b) to the melt volume (a); 
whereas, (d) is the ratio of the $1000$-year standard moving average (SMA) 
of the $\mathrm{La}$ volume (SMA of (b)) to the $1000$-year SMA of the 
melt volume (SMA of (a)).
See Section \ref{subsec:La} for physical meaning of SMA used in (d).
Grey shaded regions indicate the time interval during which the ice is 
retreating.
Different line colors correspond to different values of melt ascent 
velocity as labelled on top of the figure in $\mathrm{m/yr}$. The melt and 
$\mathrm{La}$ volumetric supply rates to the crustal chamber are the sum 
along the ridge axis
(red line in Figure \protect\ref{fig:iceMap}b)
between $45$ and $270$ $\mathrm{km}$ from the center of the ice.}
		\label{fig:volumetricRate}
	\end{center}
\end{figure}

While the \citet{Jull1996} model with constant radius of ice-load predicted 
that the region of maximum decompression rate is always below the center of 
the ice sheet (their Figure 3),
our model with variable ice radius predicts that this region is below the 
glacier terminus and is moving radially as the ice retreats. 
The glacially induced decompression causes the spatially dependent mantle 
melting rate underneath Iceland to increase from its steady state value by 
several fold during the deglaciation. These extra melts then transport to 
the surface, causing an increase in volcanic eruption rates.

The time delay between the surge of mantle melting and the surge of 
volcanic eruptions depends on the melt transport speed and also on how 
long melts reside in crustal chambers before they erupt.
Figure \ref{fig:volumetricRate}a shows the melt supply rates to crustal 
chambers predicted by our model from different input values of melt ascent 
velocity by integrating equation \eqref{eq:volumeFluxOfMelt} in the melting 
region underneath Iceland along the ridge axis from $45$ to $270$ 
$\mathrm{km}$ from the ice center, taking into account the time delay due 
to finite melt ascent velocity. The graph demonstrates that if melt 
transport were almost instantaneous, the surge in the melt supply rate 
(red curve) would respond almost immediately after the deglaciation period 
(grey shaded area). Whereas, with slower melt transport, the surge in the 
melt supply rate will be delayed from the deglaciation period. At lower 
rates of melt transport, the shape of the melt supply rate curve will be 
more
stretched in time
because melts produced at the same time at different depths 
will arrive at crustal chambers at different times.

Note that the area under the curve over the whole time interval shown in 
the graph is independent of the melt ascent velocity. This is because, by 
the conservation of mass, the total melt supply is equal to the total melt 
produced regardless of how fast the melt is transported. 

Before melts erupt on the surface, their compositions can be modified in 
the crustal chambers. We assume that the amount of melts accommodated in a 
chamber is constant. By conservation of mass, this implies that the total 
mass flux into is equal to the total mass flux out of the chamber. 
Therefore, the eruption rate is equal to the rate of melts entering the 
chamber (Figure \ref{fig:volumetricRate}a). However, the mass flux of each 
individual component do not need to follow this rule. Mixing and 
crystallization processes can modify the concentrations of REEs. We will 
discuss these two processes together with the remaining plots in Figure 
\ref{fig:volumetricRate} later in Section \ref{subsec:La}.

\newpage
\subsection{Eruptive Locations}
\label{subsec:Locations}
The ages and volumes of eruptions from the last glacial and present postglacial
are compiled using published maps and age estimates. The principal sources of
information for the Northern Volcanic Zone (NVZ) are \citet{Saemundsson1991}
and \citet{Saemundsson2012}. For the Western Volcanic Zone (WVZ) and Reykjanes
Peninsula (REYK), the maps of \citet{Sinton2005}, \citet{Eason2015} and
\citet{Saemundsson2016} are used. Acknowledgements section lists all
the sources of rock sample dataset we use in this work.

Geological mapping, geomorphology and interpretation of the volcanic 
lithologies have been used to determine the eruptive facies: whether it is 
subglacial, finiglacial or postglacial.
Finiglacial means that there is evidence of thin or recently disappeared 
ice when the eruption unit was being formed.
Finiglacial units are likely to have formed when the glacier terminus was 
sweeping through the eruptive area during the glacial retreat.

Tephrochronology provides
bounds on eruption ages for the postglacial events, meaning that the age
constraints are expressed with a maximum and minimum age bound in our dataset.
For early postglacial and finiglacial eruptions the maximum age has to be tied to
the inferred age of deglaciation of the area, based on available reconstructions
of the ice sheet history
\citep{Geirsdottir2009, Patton2017}.
The ages of subglacial eruptions are, in general, not as well constrained as those
of the postglacial. Minimum age constraints for these eruptions are obtained
from ice-sheet reconstructions and maximum ages are set to $30$ $\mathrm{ka}$.
Helium-3 exposure ages and the geomorphological characteristics of the
uppermost surface of tuyas can also be used to infer a chronology for a subset
of subglacial eruptions, using the approach of \citet{Eason2015} as informed
by the data of \citet{Licciardi2007}.

A table of eruptions for which age,
volume and chemical data is available is provided in the supplementary
information. The information in this table is used to generate the plots
provided for comparison with model results in this paper. The requirement of
an unambiguous association between sample chemistry and eruption name, volume
and age introduces some bias into our dataset: The lack of a clear link between
the eruption name and chemistry means that our coverage of subglacial eruptions
from the Reykjanes Peninsula is poor. Inevitably, erosion, superposition and
lack of subsurface informaton introduce substantial uncertainties into any
reconstruction of eruptive volumes.

\begin{figure}[ht!]
	\begin{center}
		
\includegraphics[width=1.0\textwidth]{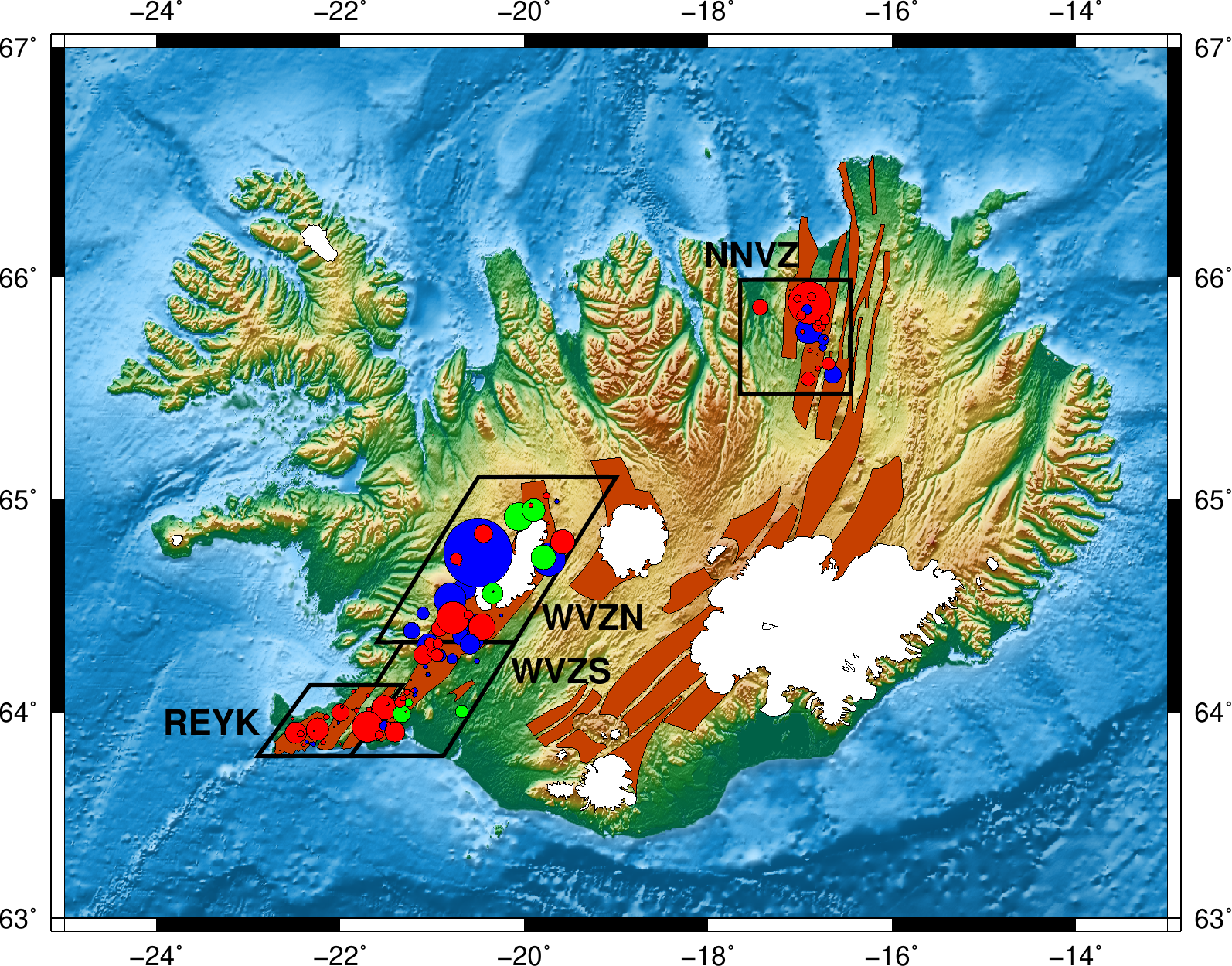}
\caption{Mercator projection map of Iceland showing locations of eruptive 
units in NNVZ, WVZN, WVZS and REYK as circles with areas proportional to 
the eruptive volumes. Colors on the circles indicate the eruption types 
(subglacial in blue, finiglacial in green and postglacial in red). See 
Section \ref{subsec:EruptionTypes} for the definition of finiglacial type. 
White areas show the recent Icelandic glaciers. Active fissure swarms 
located at where plate divergence is taking place are shown in dark red 
color. Data are provided in Supporting Information.}
		\label{fig:eruptiveLocations}
	\end{center}
\end{figure}

We divide eruption units in WVZ further into WVZ-North (WVZN) and 
WVZ-South (WVZS) by latitude of $64^{\mathrm{o}}20'0''$. Locations and types
of eruption units of all the data we use here are plotted on the map in
Figure \ref{fig:eruptiveLocations}.

The modelling-input distances of the four zones relative to the ice center 
shown in Table \ref{tab:zoneDistances} are estimates with uncertainty of 
$\approx\pm50$ $\mathrm{km}$ because the actual location of the ice center 
is unknown and also because of the uncertainty of the geometry of melt 
generation.  

\begin{table}[ht]
	\centering
		\caption{Model distances of the zones from ice center.}
		\label{tab:zoneDistances}
		\begin{tabular}{c c}
			\toprule
			\textbf{Zone}	&\textbf{Ranges (km)}\\
			\midrule
			WVZN	& $0$--$70$\\
			WVZS	& $70$--$180$\\
			NNVZ	& $120$--$180$\\
			REYK	& $180$--$250$\\
			\bottomrule
		\end{tabular}
	
\end{table}

\subsection{Eruption Types}
\label{subsec:EruptionTypes}
In this section, we show how the melt ascent velocity can affect the
relative volume proportion of different eruption types.

The modelling-input ice coverage radius as a function of time is known.
We can therefore identify if an infinitesimal volume of melt that arrives
at  the surface at a particular location and time is erupted within the
ice coverage radius or not. In other words, the model can divide eruptive 
volumes into subglacial group and subaerial group. The subglacial group 
corresponds approximately to the observational subglacial and finiglacial 
types combined. The subaerial group corresponds to the observational 
postglacial type.
Our model does not divide the subglacial group further into subglacial and 
finiglacial types.

\begin{figure}[ht!]
	\begin{center}
		
\includegraphics[width=1.0\textwidth]{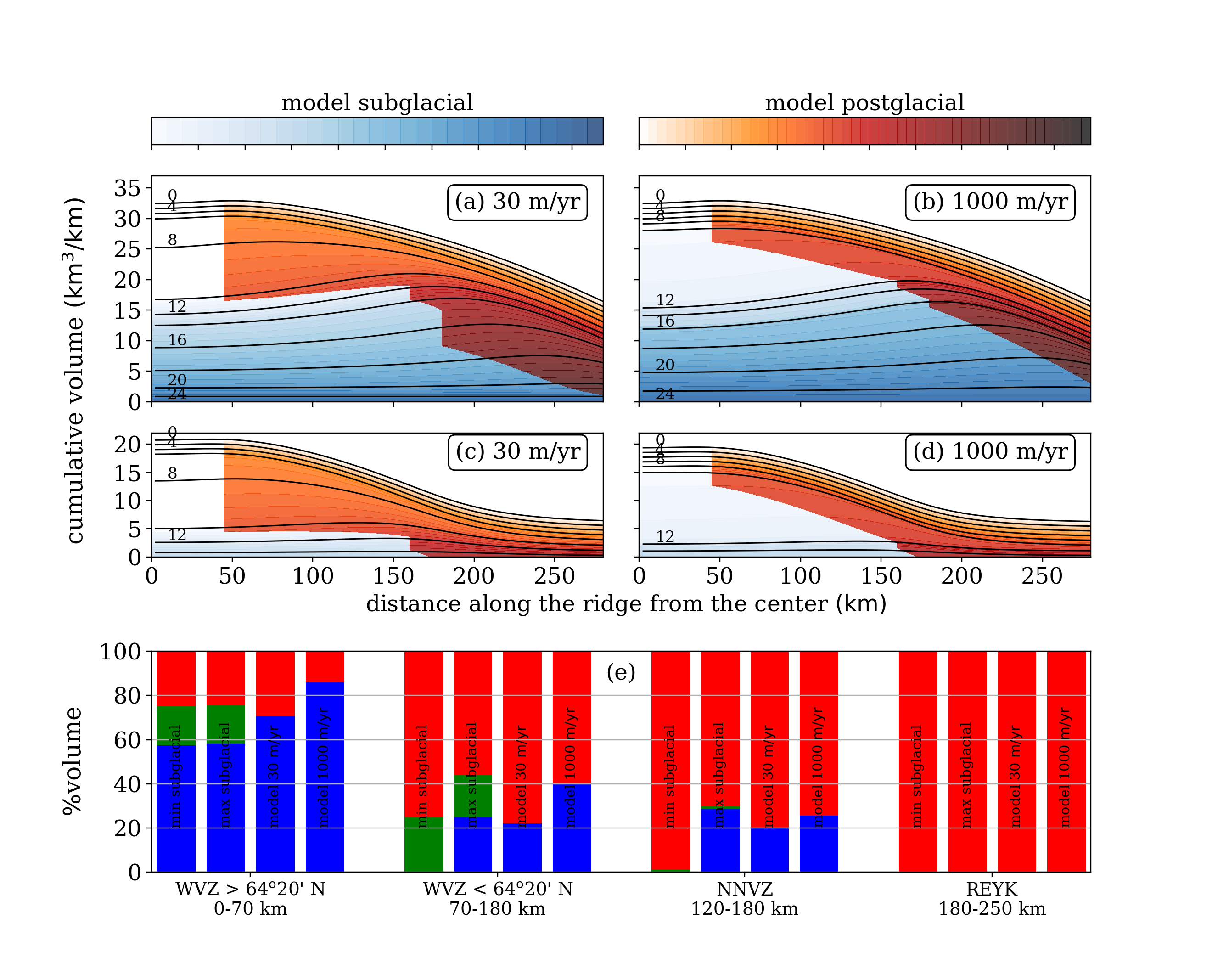}
\caption{(a)--(d): Isochrons of cumulative lava volume per unit length 
along the ridge axis as predicted by the model at melt ascent velocity of 
$30$ $\mathrm{m/yr}$ for (a) and (c) and $1,000$ $\mathrm{m/yr}$ for (b) 
and (d).
Subglacial and postglacial lavas are indicated by blue and red colors
(as indicated by the two color bars on top of the figure)
with 
color intensity proportional to the lava age.
Contour lines are separated at an equal interval of $2$ $\mathrm{kyr}$ and 
the ages labelled on the lines are in $\mathrm{kyrBP}$.
(a) and (b) show volume accumulated from $24.0$ $\mathrm{kyrBP}$; whereas, 
(c) and (d) show volume accumulated from $14.5$ $\mathrm{kyrBP}$ using the 
same modelling-inputs as in (a) and (b).
(e): Volume proportions of different eruption types that erupted between 
$14.5$ and $0$ $\mathrm{kyrBP}$ in different volcanic zones.
Observational data with lower and upper bounds of subglacial volumes are 
shown by the two left bars.
The blue, green and red bars are the subglacial, finiglacial and 
postglacial volumes respectively.
See Section \ref{subsec:EruptionTypes} for how the lower and upper bounds 
are obtained. The model results with melt ascent velocity of $30$ and 
$1,000$ $\mathrm{m/yr}$ are shown on the two right bars with blue bars 
showing the subglacial and finiglacial types combined.}
		\label{fig:lavaLayers}
	\end{center}
\end{figure}

Figure \ref{fig:lavaLayers} illustrates the model prediction that at 
faster melt transport (Figure \ref{fig:lavaLayers}b) there is a greater 
proportion of the subglacial volume
(colored blue)
compared to that at slower melt 
transport (Figure \ref{fig:lavaLayers}a). This is because faster melt 
transport will allow melts from depth to arrive at the surface sooner 
before the ice has gone.
The sharp changes of subglacial to subaerial volume at $45$, $160$ and 
$180$ $\mathrm{km}$ are due to the three pauses of the glacial terminus at 
these three radial distances (Table \ref{tab:iceTimelines}).

The model also predicts that for the same time interval (such as 
$14.5$--$0$ $\mathrm{kyrBP}$) the relative proportion of the subglacial 
volume to the total volume is dependent of the distance from the ice 
center.
This is because while most of the melts in any location are produced over 
the same time period during deglaciation ($23.0$--$10.5$ 
$\mathrm{kyrBP}$), regions closer to the ice center remain covered by ice 
for a longer period of time.
This allows a greater proportion of melts to arrive at the surface and 
erupt subglacially.
The spatial dependence of the subglacial to subaerial volume ratio is also 
seen in observations.
Figure \ref{fig:eruptiveLocations} illustrates that in the regions closer 
to the center of Iceland there is a greater proportion of subglacial and 
finiglacial volumes (blue and green circles) than further out.

To compare our model results with the observations, 
we first note that the observational eruption volumes of units older than 
$14.5$ $\mathrm{kyrBP}$ are highly uncertain not only due to glacial 
erosion but also due to some older units are buried underneath younger 
eruptions.
We therefore filter out eruptions older than $14.5$ $\mathrm{kyrBP}$ for 
both the model and the observational data.
The model cumulative volumes at melt ascent velocity of $30$ and $1,000$ 
$\mathrm{m/yr}$ after the $14.5$ $\mathrm{kyrBP}$ filter are shown on 
Figure \ref{fig:lavaLayers}c and \ref{fig:lavaLayers}d.
We use results from these two panels to calculate the model proportions of 
the subglacial volume ($=$ observational subglacial$+$finiglacial)
and subaerial volume ($=$ observational postglacial) as shown on the
two right bars of Figure \ref{fig:lavaLayers}e.
For example, the bar plot of the model $30$ $\mathrm{m/yr}$ in NNVZ on 
Figure \ref{fig:lavaLayers}e has subglacial (blue) and subaerial (red) 
proportions equal to the subglacial (blue) and subaerial (red) plotting 
area proportions of Figure \ref{fig:lavaLayers}c in the x-axis range of 
$120$--$180$ $\mathrm{km}$.

We arrange the bar plot on Figure \ref{fig:lavaLayers}e from left to right 
by zone location from the closest to (WVZN) to the furthest from (REYK) 
the ice center.
In each zone, the observational data has lower and upper estimates of 
subglacial volume due to age uncertainty of the subglacial units.
The lower estimate (min. subglacial) shown on the left bar comes from the 
volume sum of the subglacial units with maximum age bound not exceeding 
$14.5$ $\mathrm{kyrBP}$. Whereas, the upper estimate (max. subglacial) 
comes from summing all the subglacial units with minimum age bound less 
than $14.5$ $\mathrm{kyrBP}$ (while the maximum age bound can exceed 
$14.5$ $\mathrm{kyrBP}$).

The model predictions for spatial dependence in the diachronous response 
agree well with the observational data. In REYK, the whole area is already 
ice-free by $14.5$ $\mathrm{kyrBP}$ and hence all the eruptions are 
subaerial. On the other hand, WVZN remains covered by ice over most of the 
time during the last deglaciation and so the majority of the eruption 
volumes are subglacial.

Results on Figure \ref{fig:lavaLayers}e also suggests that the melt ascent 
velocity is likely to be of the order of $100$ $\mathrm{m/yr}$. At below 
$30$ $\mathrm{m/yr}$, the subglacial volumes predicted by the model would 
be smaller than that of the observational lower bound estimates (min. 
subglacial). Nevertheless, we note that the model results depend on the 
distance along the ridge axis over which the melts are integrated (as 
estimated in Section \ref{subsec:Locations}).
Similar to the model, the observational lava volumes in the four zones are
also integrated over ridge lengths of $\sim60$--$90$ $\mathrm{km}$.

\subsection{Timing of the Peaks in Volcanic Productivity}
\label{subsec:timingVol}

\begin{figure}[ht!]
	\begin{center}
		
\includegraphics[width=0.92\textwidth]{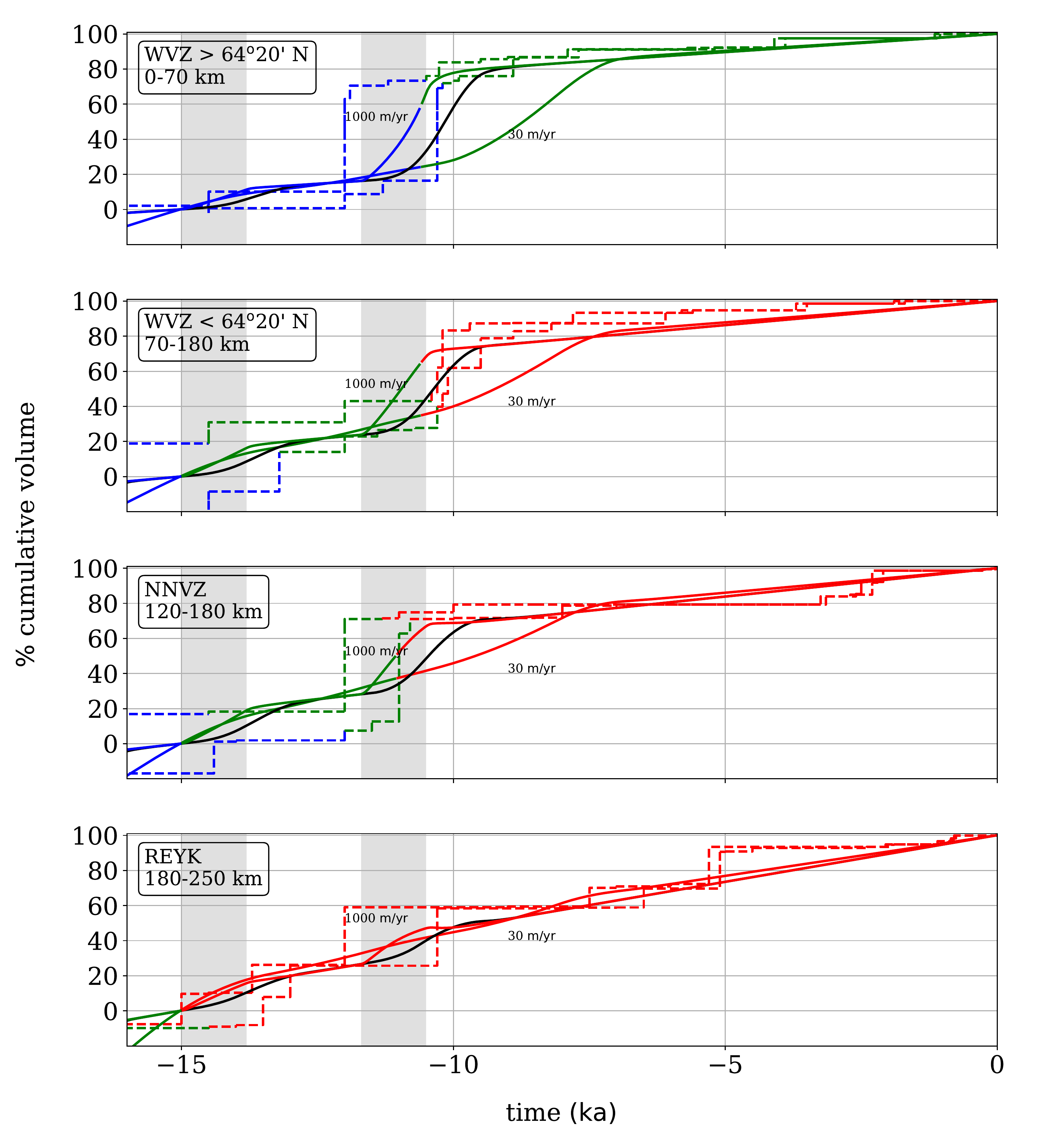}
\caption{Cumulative eruptive volume normalized to the total volume erupted 
between time $t=-15$ and $0$ $\mathrm{kyr}$.
The cumulative volumes of the observational data plotted as steps
(dashed lines) come from cumulating the eruptive volumes sorted by either
the minimum age bounds or the maximum age bounds of the eruption units.
The eruptive volume begins at $0\%$ at $-15$ $\mathrm{kyr}$ and ends at
$100\%$ at $0$ $\mathrm{kyr}$.
We use the mean cumulative volumes at these two ends to normalize the
observational data.
The model results for melt ascent velocity of $30$ and $1,000$
$\mathrm{m/yr}$ are plotted as non-black colored solid lines.
Colors on these dashed and solid lines illustrate the eruption periods: 
subglacial in blue, transitional in green and postglacial in red (see 
Section \ref{subsec:timingVol} for definition of the transitional period).
Black solid line in each panel shows the model result for melt
ascent velocity of $100$ $\mathrm{m/yr}$.
The timings of the eruption periods for the black curve are the same as
those for the remaining model-result curves.
Different panels correspond to different volcanic zones as indicated on 
the upper-left corner of each panel together with the corresponding 
modelling-input zone range (Section \ref{subsec:Locations}). Grey shaded
regions indicate the time interval during which the modelling-input ice is 
retreating.}
		\label{fig:cumvol_timing}
	\end{center}
\end{figure}

Another way to estimate the melt ascent velocity is to use the timing of 
the peaks in volcanic productivity.
On Figure \ref{fig:cumvol_timing}, we plot the cumulative eruptive volume 
as a function of time for the model outputs and the observational data. 
This figure shows that the bursts in the cumulative lava volume predicted 
by the model at melt ascent velocities between $30$ and $1,000$ 
$\mathrm{m/yr}$ have timings approximately equal to that of the 
observations across all the volcanic zones to within the uncertainties of 
the lava ages and the modelling-input ice load history.

In the period during which the glacier terminus was sweeping through each 
zone (called transitional period), some areas in the zone are already 
ice-free while some areas are still covered by ice. This means that, in 
the transitional period, eruptions can be either subglacial or subaerial.

In the observational data sorted by age, the transitional period can be 
identified approximately by the period during which there are some 
alternations of the timeline orders between subglacial, finiglacial and 
postglacial types.
The remaining two end periods are called subglacial and postglacial 
periods.
The subglacial
period consists of only subglacial type and
the
postglacial period consists of only postglacial type. In the model, the 
transitional period is identified by the period during which the ice radius 
is in the zone range (as listed in Table \ref{tab:zoneDistances}).

The timing of the periods in the model is controlled by the input ice 
history and the input zone range. Therefore, a well-matched timing of the 
periods between the model and the observational data helps us identify if 
the modelling-input of ice load and zone range reasonably reflect the 
actual values.
The discrepancy between the observations and the model results at our
preferred estimate of $100$ $\mathrm{m/yr}$ melt ascent velocity is likely
to be because of the uncertainty of the observational eruption ages and
the model axisymmetric ice assumption.

Nevertheless, the results in Figure \ref{fig:cumvol_timing}
show that,
at a melt 
ascent velocity of $30$ $\mathrm{m/yr}$ or below, the difference of the 
timing in the burst in volcanism between the model results and the 
observational data becomes significant.
More quantitatively, with the average mantle melting depth of $\approx50$ 
$\mathrm{km}$, reducing the melt ascent velocity from $100$ to $30$ 
$\mathrm{m/yr}$ will delay the burst timing by $\approx1.2$ 
$\mathrm{kyr}$, which is significant compared to the uncertainty of the 
eruption ages.
Our model implies
a
similar lower bound value to that of $50$ 
$\mathrm{m/yr}$ estimated in \citet{Maclennan2002} from the relative timing 
between the observed eruption ages and the deglaciation.
An advantage of this work is that much broader geographical spread is 
accounted for. Our model
examines
eruptions in different volcanic zones; 
whereas, \citet{Maclennan2002} only examined eruptions in the NNVZ region.

\newpage
\subsection{Geochemical Response}
\label{subsec:La}

\subsubsection{Model Predictions}
\label{subsubsec:La_ModelResults}
Figure \ref{fig:volumetricRate}b shows the volumetric supply rate of 
$\mathrm{La}$ to the crustal chambers normalized to the steady-state 
$\mathrm{La}$ concentration. Similarly to the whole melts in Figure 
\ref{fig:volumetricRate}a, the surge in the supply rate of $\mathrm{La}$ 
is delayed from the deglaciation period due to the finite speed of melt 
transport.
However,
while the volumetric melt supply rate curves
(Figure \protect\ref{fig:volumetricRate}a)
are stretched in time, the $\mathrm{La}$ supply rate curves
(Figure \protect\ref{fig:volumetricRate}b)
retain their shapes almost like the same time-series but time-shifted.
This is because $\mathrm{La}$ is partitioned into melt at almost the same
depth (near the solidus).
Most of the $\mathrm{La}$ takes almost an equal time to arrive at
the crust regardless of how fast the ascent rate is.
Therefore, changing the ascent rate will not significantly spread the
$\mathrm{La}$ flux out along the time axis.
In contrast, melts are produced at different depths.
They take different times to transport to the crust.
The slower the ascent rate, the more the time delay between melts from
different depths to arrive at the surface; hence, the more the spread
of the melt supply rate curve along the time axis.

Figure \ref{fig:volumetricRate}c is the $\mathrm{La}$ concentration 
($c_{\mathrm{La}}$) in the melt supply to the crustal chambers, which is 
equal to Figure \ref{fig:volumetricRate}b divided by Figure 
\ref{fig:volumetricRate}a. This would correspond to the concentration in 
the lava erupted at the surface if the magma mixing process in the crustal 
chamber were not present. The longer the magma is allowed to mix in the 
crustal chamber, the smaller the variation signal of the REE 
concentrations.

In the model, the effect of magma mixing on $c_{\mathrm{La}}$ in the lavas 
is equivalent mathematically to the time average concentration. The time 
period over which the average is performed is equal to the time duration 
that the magmas mix in the crustal chamber before they erupt. In a study 
of the chemical disequilibria between olivine, its melt-inclusions and the 
whole melts that surround the olivine in rock samples collected from 
Iceland, \citet{Maclennan2008} estimated that the magma residence time is 
of the order of a few hundreds to $\lesssim 1,000$ years. We therefore 
take the time average $c_{\mathrm{La}}$ over a period of $1,000$ years and 
the result is shown in Figure \ref{fig:volumetricRate}d. In other words, 
this panel is equal to the ratio between the $1000$-year standard moving 
average (SMA) of $\mathrm{La}$ volume supply to the crustal chamber 
($1000$-year SMA of Figure \ref{fig:volumetricRate}b) and the $1000$-year
SMA of the whole melt volume supply to the crustal chamber ($1000$-year
SMA of Figure \ref{fig:volumetricRate}a).

Note that melt mixing also occurs "en-route" while melts are migrating 
from depths to the crustal chamber and meet together along the way. This 
geological process corresponds mathematically to the volume integration 
over the mantle melting domain $\mathcal{V}$ as shown in
equation \eqref{eq:volumeFluxOfElement},
taking into account the time delay due to the 
finite rate of melt transport $\Delta t$.
In other words,
the model results shown in Figures \protect\ref{fig:volumetricRate}b,
\protect\ref{fig:volumetricRate}c and \protect\ref{fig:volumetricRate}d
as calculated by equations \protect\eqref{eq:volumeFluxOfElement} and
\protect\eqref{eq:concentrationFluxOfElement}
have already taken into account the effect of en-route melt mixing.

Our results in Figure \ref{fig:volumetricRate}c and 
\ref{fig:volumetricRate}d show that the variation of $c_{\mathrm{La}}$ is 
strongly dependent on the melt ascent velocity.
The lower the melt ascent velocity the higher the variation of 
$c_{\mathrm{La}}$. This effect can be explained as follows.
During the deglaciation, the decompression rate in the mantle is maximum 
at the surface and decays exponentially with depth (as illustrated in 
Figure \ref{fig:contourDpDt}). This means that the extra melts generated 
during deglaciation are mostly produced at shallow depths in the mantle, 
which is depleted in $\mathrm{La}$. If the melt transport had been 
instantaneous, the extra melts produced at any depth at the same time would 
have travelled to the crust and mixed instantly and would have erupted at 
the surface with $\mathrm{La}$ depletion of up to $\approx 20\%$ as 
predicted by \citet{Jull1996}. In contrast, when the melt ascent velocity is 
finite, the extra melts produced at shallower depths during deglaciation 
will arrive at the surface before the extra melts produced at deeper 
depths. The slower the rate of melt transport the more likely the extra 
melts from shallow depths ($\mathrm{La}$ depleted) are to erupt before they 
mix with the extra melts from deep depths ($\mathrm{La}$ enriched). As a 
result, 
when the melt ascent velocity is sufficiently low, the first arrival of 
the extra melts produced during deglaciation will be much more depleted in 
$\mathrm{La}$ than that predicted by the instantaneous melt transport 
model of \citet{Jull1996}.

The eruptive $c_{\mathrm{La}}$ will recover back to near the steady-state 
concentration after the extra melts from the bottom of the melting region 
($\mathrm{La}$ enriched) catch up and mix with the extra melts from 
shallow depths ($\mathrm{La}$ depleted) before they erupt.
Moreover, the recovery of the eruptive $c_{\mathrm{La}}$ back to the 
steady-state will overshoot after the deglaciation ends.
This phenomenon can be explained as follows. Once the deglaciation 
terminates, the glacially-induced decompression melting in the mantle will 
also terminate at all depths at the same time and the extra melt supply to 
the surface from shallow depths ($\mathrm{La}$ depleted) will run out 
before the extra melt supply from deep depths ($\mathrm{La}$ enriched).
This is because the melts from greater depths take a longer time to arrive 
at the surface.
Therefore, once the $\mathrm{La}$ depleted melt supply from the shallow 
depths runs out, the remaining majority of the erupted lavas will be the 
$\mathrm{La}$ enriched melts from deep depths and the eruption will become 
enriched in $\mathrm{La}$.

Figure \ref{fig:volumetricRate}d also shows that the timing of the periods 
during which the lavas are enriched or depleted in $\mathrm{La}$ is 
dependent on the melt ascent velocity.
At slower melt ascent velocity, the peaks and the troughs of 
$c_{\mathrm{La}}$ are delayed further from the deglaciation periods.
Hence, we can use this timing combined with the magnitude of the 
$c_{\mathrm{La}}$ variations to estimate the melt ascent velocity.
We note that the $\mathrm{La}$ depleted lava volume dominates the 
$\mathrm{La}$ enriched lava volume.
This can be seen in Figure \ref{fig:volumetricRate}.
The troughs of $c_{\mathrm{La}}$ (Figure \ref{fig:volumetricRate}d) fall 
in the periods of the bursts in eruption rates (Figure 
\ref{fig:volumetricRate}a); whereas, the peaks of $c_{\mathrm{La}}$ 
(Figure \ref{fig:volumetricRate}d) fall outside those periods.
Therefore, the $\mathrm{La}$-depletion signal is stronger than the 
$\mathrm{La}$-enrichment signal, which is also seen in observational data.
The majority of the eruptions during the last deglaciation are depleted in 
$\mathrm{La}$.

\newpage
\subsubsection{Geological Observations}
\label{subsubsec:La_Observations}
The eruptive $\mathrm{La}$ concentrations of observational data are from 
rock samples collected from Iceland by the previous studies (see
Acknowledgements section and Supporting Information for details).
These rock samples are of melts that have gone through 
fractionation/accumulation during cooling and crystallization processes in 
the crustal chambers.
These processes modify the melt compositions from their original 
pre-crustal compositions.

We make fractionation/accumulation correction of $c_{\mathrm{La}}$ in each 
rock sample based on the $\mathrm{MgO}$ content of the sample.
We assume that the pre-crustal melts have $14.0$ $\mathrm{wt\%}$ 
$\mathrm{MgO}$ as estimated in \citet{Maclennan2001}.
Rock samples that have $\mathrm{MgO}$ between $9.5$ and $14.0$ 
$\mathrm{wt\%}$ are assumed to have undergone crystallization of 
olivine-rich material with $40.0$ $\mathrm{wt\%}$ $\mathrm{MgO}$.
If the melts underwent
crystallization
further below $9.5$ $\mathrm{wt\%}$ 
$\mathrm{MgO}$, they generate a gabbroic solid with $11.0$ $\mathrm{wt\%}$ 
$\mathrm{MgO}$. Some rock samples have higher $\mathrm{MgO}$ content than 
$14.0$ $\mathrm{wt\%}$ of the pre-crustal melts.
We assume that these samples are from melts that have been influenced by 
the accumulation of olivine crystals with $40.0$ $\mathrm{wt\%}$ 
$\mathrm{MgO}$.

\begin{figure}[ht!]
	\begin{center}
		\includegraphics[width=1.0\textwidth]{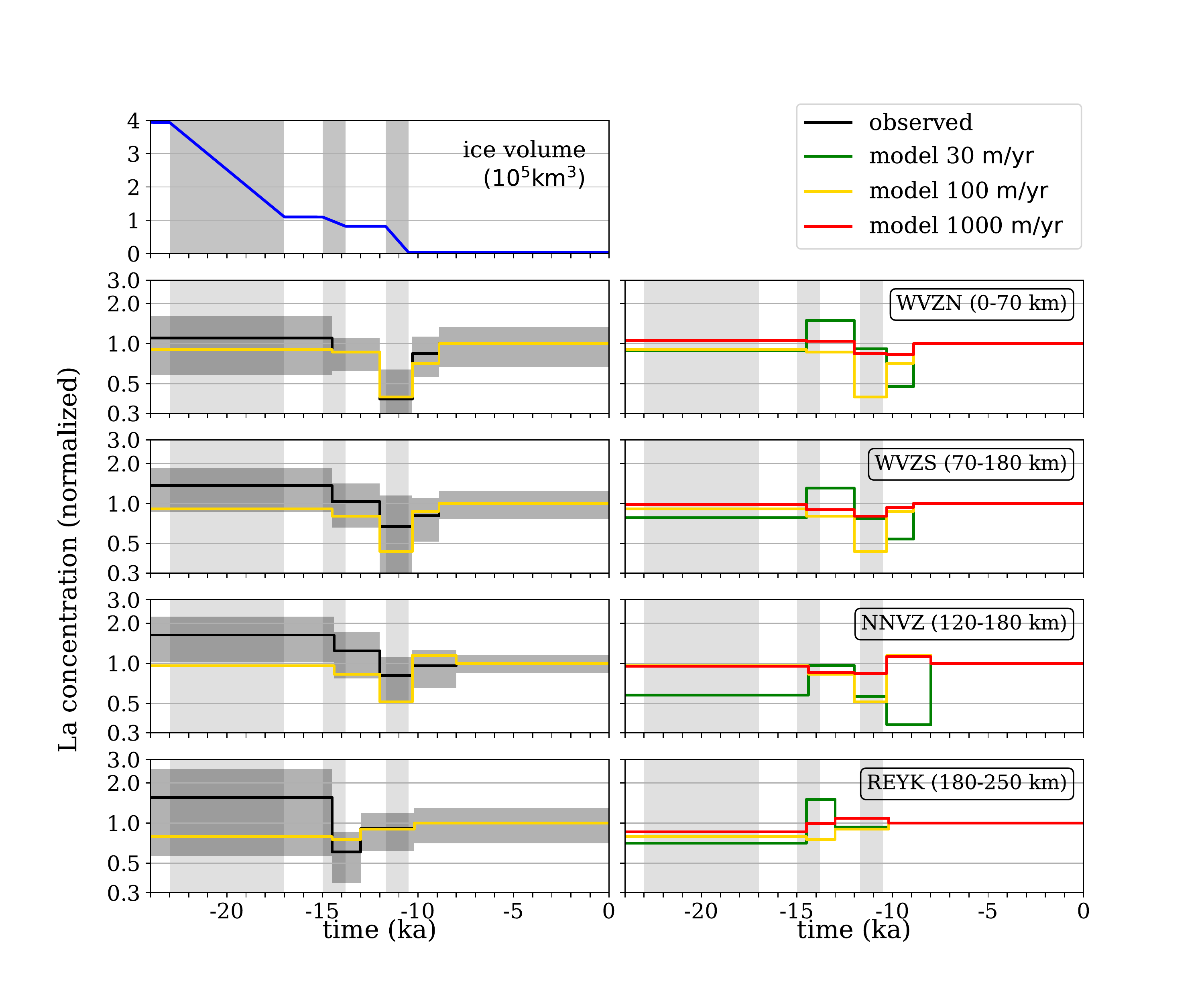}
\caption{Top panel is the modelling-input ice volume. Each row of the 
remaining panels shows $\mathrm{La}$ concentrations normalized to the 
steady-state concentration in each volcanic zone. The upper-right corner 
of the right panel labels the corresponding zone and the modelling-input 
zone range.
On the left panel, the black curve shows the observational $\mathrm{La}$ 
concentrations together with grey bands indicating $\pm1$ S.D. of rock 
samples. The yellow curve is the model result from a melt ascent velocity 
of $100$ $\mathrm{m/yr}$.
The right panel illustrates the model $\mathrm{La}$ concentrations 
calculated from different values of melt ascent velocity labelled with 
different line colors.
Details of how the $\mathrm{La}$ concentrations are calculated can be 
found in the text.}
		\label{fig:La}
	\end{center}
\end{figure}

Due to the age uncertainty of eruption units, $c_{\mathrm{La}}$ cannot be 
plotted directly as that of the model in Figure \ref{fig:volumetricRate}.
Most of the eruptions have their age estimated as a time band bounded by 
some geological events with identifiable age (e.g. tephra layers).
In WVZ and NNVZ, most of the eruption units fall into one of the following 
age bands:
\begin{enumerate}
\item[1.]
Glacial (pre $14.5$/$14.4$ $\mathrm{kyrBP}$)
\item[2.]
Eruptive Pulse 1 ($14.5$/$14.4$ to $12.0$ $\mathrm{kyrBP}$)
\item[3.]
Eruptive Pulse 2 ($12.0$ to $10.3$ $\mathrm{kyrBP}$)
\item[4.]
Early Postglacial ($10.3$ to $8.9$/$8.0$ $\mathrm{kyrBP}$)
\item[5.]
Steady-State Postglacial (post $8.9$/$8.0$ $\mathrm{kyrBP}$)
\end{enumerate}
REYK zone is different in that it is the furthest from the ice center and 
became ice-free by $14.5$ $\mathrm{kyrBP}$, which is earlier than the 
other zones.
Eruption units in REYK are divided into the following age bands:
\begin{enumerate}
\item[1.]
Glacial (pre $14.5$ $\mathrm{kyrBP}$)
\item[2.]
Early Postglacial 1 ($14.5$ to $13.0$ $\mathrm{kyrBP}$)
\item[3.]
Early Postglacial 2 ($13.0$ to $10.2$ $\mathrm{kyrBP}$)
\item[4.]
Steady-State Postglacial (post $10.2$ $\mathrm{kyrBP}$)
\end{enumerate}
In each of these age bands, we take the volume-weighted average 
$\mathrm{La}$ concentration normalized to the Steady-State Postglacial 
$\mathrm{La}$ concentration and plot in Figure \ref{fig:La} for both the 
observational data and the model.

The right column of Figure \ref{fig:La} illustrates that different model 
melt ascent velocities result in different $\mathrm{La}$ concentration 
characteristics.
In each age band, some values of melt ascent velocity may predict 
$\mathrm{La}$ depletion, whereas the others predict $\mathrm{La}$ 
enrichment.
This is due to the effect of melt ascent velocity on the timing of the 
peaks and troughs of $c_{\mathrm{La}}$ (Figure \ref{fig:volumetricRate}d) 
that we discussed earlier.

The left column of Figure \ref{fig:La} shows that the model melt ascent 
velocity of $100$ $\mathrm{m/yr}$ yields similar $c_{\mathrm{La}}$ 
characteristics to that of the observations.
At two extreme melt ascent velocities of $30$ and $1,000$ $\mathrm{m/yr}$, 
the $c_{\mathrm{La}}$ characteristics are significantly different from 
that of the observations both in the timing and the magnitude of the 
$c_{\mathrm{La}}$ variations.

In the Glacial age band, the observational $c_{\mathrm{La}}$ is elevated 
from the Steady-State Postglacial value in all the volcanic zones.
This is likely to be due to the effect of glacial loading on 
depth-dependent melting suppression that occurred before the Last Glacial 
Maximum ($23$ $\mathrm{kyrBP}$).
This feature is not included in our model here, which may explain why the 
model $c_{\mathrm{La}}$ in the Glacial age band is lower than that 
observed across all the volcanic zones.
One of our future works will be to investigate this glacial loading effect.

\subsection{Model Limitations}
The accuracy of our results depends on several factors.
The deviations of modelling input parameters from the actual geological 
values that are not well-constrained can be significant.

For example, the model $\mathrm{La}$ concentration is dependent on the 
time period over which the magma mixes in the crustal chamber. As mentioned 
in Section \ref{subsec:La}, the longer the magma residence time, the lower 
the variations of $\mathrm{La}$ concentrations.
Also, the residence time may not be the same throughout Iceland as assumed
in our model.
A better constraint on the effective magma residence time in the chamber
may therefore be required.

Our $~100$ $\mathrm{m/yr}$ estimate of the melt ascent velocity likely
represents that of the melt produced during the GIA.
At steady state, the decompression melting rate is significantly less.
This leads to a significantly lower mass flux of melt and likely results
in a slower rate of melt transport.
This could be one reason why our ascent rate estimate is significantly
higher than that in models of melt transport at Mid-Ocean Ridges 
\citep{Burley2015, Crowley2015}.
A melt ascent velocity of $100$ $\mathrm{m/yr}$ is $\sim2$ orders of
magnitude faster than would be predicted from simple models of diffuse
porous flow.
As has been noted in several previous studies, such rapid melt ascent
velocities suggest that some focusing or channelization of melt must
occur during transport (e.g. \cite{Kelemen1997}).

In more elaborate fluid dynamic models (e.g. \protect\cite{McKenzie1984}),
the melt velocity varies with depth. 
Melt flow starts slow at the base of the melting region and ascends at a
faster rate as it migrates to a shallower depth where the porosity is
higher.
Therefore, in any region below the GIA average melting depth, the melt
ascent velocity is likely to be below our estimate.
This depth-dependent melt ascent velocity would cause a more time delay
of the burst of the $\mathrm{La}$ supply rate to the crustal chamber than
that predicted by our model in Figure \ref{fig:volumetricRate}b.
A larger time delay between the melt supply and the $\mathrm{La}$ supply
would increase the time intervals during which the $\mathrm{La}$
concentration (Figures \ref{fig:volumetricRate}c and
\ref{fig:volumetricRate}d) is depleted or enriched.

2-D fluid dynamic models of melt and trace element transport
(e.g. \protect\cite{Spiegelman1996}) also predict an across-axis variation
in the erupted melt composition.
In our model, we assume complete melt mixing and extraction on the
ridge axis.
This produces only a single average concentration of $\mathrm{La}$ at each
snapshot in time.
\protect\cite{Spiegelman1996} also showed that
the convergence of melt to the ridge axis in passive ridge flow leads to
an enrichment of incompatible elements in the erupted melt by almost a
factor of $2$ (for $D^i\leq0.01$) from that in the 1-D column model.
If the full solution of melt transport had been incorporated into our
model, the $\mathrm{La}$ concentration at steady state would have also
been increased by a factor of $\sim2$.
If the same enrichment factor ($\sim2$) also uniformly applies to that
during the GIA, our model results of the $\mathrm{La}$ concentration
(normalized to the steady state value) would remain unchanged.
However, the flow fields of the solid mantle and of the melts during the
GIA are certainly different from those at steady state.
Therefore, the enrichment factor during the GIA is unlikely to be
uniformly the same as that at the steady state.
How much the enrichment factor varies still remains to be explored.
In our future work, we would like to incorporate full melt transport
solutions into the model to understand how good the constant ascent rate
approximation is.

The real ice sheet shape may be significantly deviated from the 
axisymmetric shape that we use. While our axisymmetric assumption helps 
simplify the computations, a modelling-input ice sheet with more detailed 
3D shape may
have
an important role in controlling the accuracy of the 
model results. 

Last but not least, the time evolution of the ice sheet shape we input 
into our model may be significantly different from the actual ice sheet. 
For example, the glacier may extensively re-advance in some periods during 
the last deglaciation. Glacial advance will increase the load on the 
surface, which will lead to pressure increase in the mantle. This will 
suppress mantle melting and can also affect the REE concentrations as 
discussed in Section \ref{subsec:La}. The modelling of the effects of 
glacial advance on mantle melting beneath Iceland may therefore be 
important.

\section{Conclusions}
The consequences of a finite melt ascent velocity on lavas erupted during 
the last deglaciation are:
\begin{enumerate}
\item[1.]
Volume proportions of different eruption types:
Faster melt transport will allow more melts to arrive at the surface and 
erupt sooner when the ice is still present. This means that there will be 
a greater proportion of subglacial and finiglacial volumes relative to 
postglacial volume.
\item[2.]
Relative timing between the bursts in the eruption rates and the 
deglaciation:
Higher melt ascent velocity will transport the extra melts produced during 
deglaciation to the surface faster. This will result in
a
smaller time-lag 
between the bursts in the eruption rates and the deglaciation.
\item[3.]
Variations of REE concentrations:
Slower melt ascent velocity will result in a greater time-lag between 
melts from shallow depth (REE depleted) and melts from deep depth (REE 
enriched) arriving at the surface. This will cause higher variations of 
REE concentrations in the lavas.

\end{enumerate}

Our numerical model estimates that the Icelandic melt transport from the 
upper mantle melting region to the surface during the last-deglaciation 
has an average melt ascent velocity of the order of
$\sim100$ $\mathrm{m/yr}$.

\newpage
\appendix
\section{Mantle Flow}
\label{appendix:mantleFlow}
\setcounter{equation}{0}

In steady state, the velocity components of the mantle flow follow the 
corner flow solutions. In Cartesian coordinates, they can be written as 
(\cite{Batchelor2000}; \cite{Spiegelman1987})

\begin{linenomath*}
\begin{align}
	v_{x}(x,y,z) & = \frac{B\,x\,z}{x^2 + z^2} - 
B\,\mathrm{arctan}\left(\frac{x}{z}\right), \nonumber \\
	v_{y}(x,y,z) & = 0, \nonumber\\
	\textrm{and} \qquad
	v_{z}(x,y,z) & = \frac{B\,z^2}{x^2 + z^2} - 
B\,\mathrm{sin^2}(\alpha) \label{eq:cornerFlow}
\end{align}
\end{linenomath*}
where
\begin{linenomath*}
\begin{align*}
	B & = \frac{2\,U_{0}}{\pi - 2\alpha - \mathrm{sin}(2\alpha)} \, ,
\end{align*}
\end{linenomath*}
$v_{x}$ is the horizontal velocity component perpendicular to the ridge, 
$v_{y}$ is the horizontal velocity component parallel to the ridge, 
$v_{z}$ is the velocity component in the vertical direction, $U_{0}$ is 
the half-spreading velocity of the ridge and $\alpha$ is the ridge angle 
from the horizontal.

Ice load change causes glacially-induced isostatic adjustment (GIA). In 
cylindrical coordinates, the semi-analytical solutions to the 
axi-symmetric GIA response in the viscous half-space mantle are

\begin{linenomath*}
\begin{align}
	v_r(r,z,t) = & - \frac{\rho_i}{\rho_s} \, \mathcal{H}^{-1}_{1}
	\left[
		kz\,e^{kz}\,
		\widetilde{w}(k,t)
	\right]
	\, , \nonumber
	\\
	v_{\theta} (r,z,t) = & \,0
	\, , \nonumber
	\\
	v_z(r,z,t) = & - \frac{\rho_i}{\rho_s} \, \mathcal{H}^{-1}_{0}
	\left[
		(1-kz)\,e^{kz}\,
		\widetilde{w}(k,t)
	\right]
	\, , \nonumber
	\\
	P\,(r,z,t) = & -\rho_s gz
	+\, \rho_i g \, \mathcal{H}^{-1}_{0}
	\left[
		\tau \,e^{kz}\,
		\widetilde{w}(k,t)
	\right]
	\, , \nonumber
	\\
	\textrm{and} \qquad
	\frac{DP}{Dt} \,(r,z,t) = & \,
	\rho_i g \, \mathcal{H}^{-1}_{0}
	\left[ e^{kz}\,
		\left(
			\mathcal{H}_{0}[\dot{h}](k,t) \,
			-kz \, \widetilde{w}(k,t)
		\right)
	\right],
	\label{eq:rebound}
\end{align}
\end{linenomath*}
where
\begin{linenomath*}
\begin{align*}
	\tau \equiv & \,\tau(k) = \frac{2\,\eta\,k}{\rho_s\,g}
	\, , \nonumber \\
	\widetilde{w}\,(k,t) = & \int_{t_{\mathrm{iso}}}^t
	\left( \mathcal{H}_{0}[\dot{h}](k,t') \right)\,\,
	\mathrm{exp}\left(-\frac{t-t'}{\tau}\right)\,
	\mathrm{d}t'/\tau
	\, , \nonumber \\
	\mathcal{H}_{n}[f](k) = & \int_0^{\infty}
	f(r)\, \mathrm{J}_n (kr) \, r\, \mathrm{d}r
	\, , \nonumber \\
	\mathcal{H}^{-1}_{n}[F](r) = & \int_0^{\infty}
	F(k)\, \mathrm{J}_n (kr) \, k\, \mathrm{d}k
	\, , \nonumber
\end{align*}
\end{linenomath*}
$v_{r}$ is the radial component of the velocity, $v_{\theta}$ is the 
azimuthal component of the velocity, $v_{z}$ is the vertical component of 
the velocity, $P$ is the pressure in the mantle, $\rho_i$ is the density 
of ice, $\rho_s$ is the density of the mantle, $\dot{h}$ is the 
time-derivative of the thickness of ice sheet, $\mathcal{H}_n[f]$ is the 
$n^{\text{th}}$-order Hankel transform of function $f$, 
$\mathcal{H}^{-1}_n[F]$ is the $n^{\text{th}}$-order inverse Hankel 
transform of function $F$,
$\mathrm{J}_n$ is the $n^{\text{th}}$-order 
Bessel function of the first kind
and $k$ is the wavenumber.
$t_{\mathrm{iso}}$ in the 
$\widetilde{w}\,(k,t)$ expression is the time at which the mantle is in 
isostatic equilibrium.

Equation \eqref{eq:rebound} shows that the glacially induced decompression 
rate depends on the history of the deglaciation rate $\dot{h}$ and is 
attenuated exponentially with depth by the $e^{kz}$ factor ($z < 0$ in the 
mantle).
For the ice sheet shape in equation \eqref{eq:iceProfile}, the Hankel 
transform of the rate of change of ice load is analytical:
\begin{linenomath*}
\begin{equation}
\mathcal{H}_{0}[\dot{h}](k,t) =
\frac{3\sqrt{2}}{32} \, \dot{V}(t)
\left[
\mathrm{J}_{-\frac{1}{4}}\left(\frac{k\, r_m(t)}{2}\right) \, 
\mathrm{J}_{\frac{1}{4}}\left(\frac{k\, r_m(t)}{2}\right)
+
\mathrm{J}_{-\frac{3}{4}}\left(\frac{k\, r_m(t)}{2}\right) \, 
\mathrm{J}_{\frac{3}{4}}\left(\frac{k\, r_m(t)}{2}\right)
\right].
\label{eq:iceHankel}
\end{equation}
\end{linenomath*}

\section{Numerical Methods}
\label{appendix:numericalMethods}
\setcounter{equation}{0}

Calculations of the melting rates, the eruption rates and the REE 
concentrations (equations \eqref{eq:massFluxOfMelt},
\eqref{eq:volumeFluxOfMelt}, \eqref{eq:massFluxOfElement} and
\eqref{eq:volumeFluxOfElement})
require temporal and spatial integrations over a finite domain.
We perform these numerical integrations using the trapezoidal rule.
We discretize the spatial domain using a 3-D rectangular grid with 
uniform horizontal and vertical resolutions of $5$-by-$5$ $\mathrm{km^2}$ 
and $0.5$ $\mathrm{km}$ respectively.

For an REE with $D^i\ll1$, the concentration $c^i_l$ changes rapidly with 
depth near the solidus.
As can be seen in equation \eqref{eq:instantaneousConcentration}, $c^i_l$ 
drops sharply with $F$ near the solidus $F=0$.
By adopting the trapezoidal rule to integrate equation
\eqref{eq:massFluxOfElement} along the depth using values of $c^i_l$ at the
grid vertices alone, the trapezoidal error can be very significant.
In order to resolve this rapid change within each cell of the grid, we 
calculate $c^i_l$ inside the cell using equation
\eqref{eq:instantaneousConcentration} with $F=F(z)$ that is obtained from 
linear interpolation of the face-averaged $F$ between the lower face and 
upper face of the cell with depth $z$.
The face-averaged value of a face is simply the mean value of the four 
vertices at the corners of the face.
This technique helps improve the model $c^i_l$ accuracy significantly.

Calculations of mantle flow and decompression rates such as equation
\eqref{eq:rebound} involve the inverse Hankel transform, which requires 
numerical integration of the wavenumber $k$ from $0$ to $\infty$.
All the mathematical expressions in the model that require inverse Hankel 
transform contain an attenuation factor $e^{kz}$, which decays 
exponentially with the wavenumber $k$ in the mantle ($z<0$).
Therefore, the numerical integration of the inverse Hankel transform from 
$k=0$ to $\infty$ can be truncated when the attenuation factor $e^{kz}$ is 
negligibly small.
In our model, the melting region is at $z=-20$ $\mathrm{km}$ or below. We 
truncate the integration at $k=2/3$ $\mathrm{km}^{-1}$, which corresponds 
to $e^{kz}\sim2\times10^{-6}$ or below.

The variations of integrands of all the inverse Hankel transform involved 
in the model are dominated by the ice load function in the $k$-domain 
(equation \eqref{eq:iceHankel}).
This function consists of the $n^{\text{th}}$-order Bessel functions of 
the first kind  $\mathrm{J}_n$ with $n=\pm1/4$ and $\pm3/4$, all of which 
have the same argument $=kr_m/2$ where $r_m$ is the ice radius.
This means that the ice load function in the $k$-domain varies with $k$ at 
a frequency of $\sim r_m/2 \sim100$ $\mathrm{km}$.
We therefore use the trapezoidal strip size $\mathrm{d}k=1/1440$ 
$\mathrm{km}^{-1}$, which gives $\mathrm{d}kr_m/2\leq10^{-1}$.
This corresponds to having at least $N_k=10$ trapezoidal strips per unit 
length in the non-dimensional $k$-domain since 
$N_k=2/(\mathrm{d}kr_m)\geq10$.

The
decompression rate $=-DP/Dt$
at time $t$ can be calculated directly from 
equation \eqref{eq:rebound} independently from any information in the 
previous time steps.
This means that the time-step size does not affect the accuracy of the 
model. We use a uniform time-step size of $50$ $\mathrm{yr}$.

\section*{Acknowledgements}
We would like to thank Marc W. Spiegelman, Deborah Eason and an
anonymous reviewer for their constructive and thorough reviews, which greatly
helped improve this manuscript. 
This research
is funded by
the Cambridge Trust and the Leverhulme Trust.
All the data of rock samples we use (as provided in the Supporting Information)
come from the following published sources:
\citet{Brandon2000}; 
\citet{Brandon2007}; \citet{Breddam2000}; \citet{Burnard2005}; 
\citet{Chauvel2000}; \citet{Condomines1983}; \citet{Debaille2009}; 
\citet{Dixon2000}; \citet{Dixon2003}; \citet{Eason2009}; \citet{Eason2015}; 
\citet{Elliott1991}; \citet{Fitton2003}; \citet{Furi2010}; \citet{Gee1998a}; 
\citet{Gee1998b}; \citet{Geirsdottir2009};
\citet{Hardarson1997}; \citet{Hemond1993}; 
\citet{Jakobsson1978}; \citet{Jonasson2005}; \citet{Kempton2000}; 
\citet{Kokfelt2006}; \citet{Koornneef2012}; \citet{Kurz1985}; 
\citet{Maclennan2001}; \citet{Maclennan2003}; \citet{Maclennan2004}; 
\citet{Macpherson2005}; \citet{Nicholson1991}; \citet{Nielsen2007}; 
\citet{Peate2009}; \citet{Peate2010}; \citet{Poreda1986};
\citet{Saemundsson1991}; \citet{Saemundsson2012}; \citet{Saemundsson2016};
\citet{Sigurdsson1978}; \citet{Sinton2005}; \citet{Skovgaard2001}; 
\citet{Slater2001}; \citet{Sobolev2008}; \citet{solnes2013}; 
\citet{Stracke2003}; \citet{Thirlwall2004}; \citet{
Thirlwall2006}


%

\begin{thebibliography}{}

\bibitem [\protect \citeauthoryear {%
{\'A}rnad{\'o}ttir%
\ \protect \BOthers {.}}{%
{\'A}rnad{\'o}ttir%
\ \protect \BOthers {.}}{%
{\protect \APACyear {2009}}%
}]{%
Arnadottir2009}
\APACinsertmetastar {%
Arnadottir2009}%
\begin{APACrefauthors}%
{\'A}rnad{\'o}ttir, T.%
, Lund, B.%
, Jiang, W.%
, Geirsson, H.%
, Bj{\"o}rnsson, H.%
, Einarsson, P.%
\BCBL {}\ \BBA {} Sigurdsson, T.%
\end{APACrefauthors}%
\unskip\
\newblock
\APACrefYearMonthDay{2009}{}{}.
\newblock
{\BBOQ}\APACrefatitle {{Glacial rebound and plate spreading: results from the
  first countrywide GPS observations in Iceland}} {{Glacial rebound and plate
  spreading: results from the first countrywide GPS observations in
  Iceland}}.{\BBCQ}
\newblock
\APACjournalVolNumPages{Geophysical Journal International}{177}{2}{691--716}.
\newblock
\begin{APACrefDOI} \doi{10.1111/j.1365-246X.2008.04059.x} \end{APACrefDOI}
\PrintBackRefs{\CurrentBib}

\bibitem [\protect \citeauthoryear {%
Batchelor%
}{%
Batchelor%
}{%
{\protect \APACyear {2000}}%
}]{%
Batchelor2000}
\APACinsertmetastar {%
Batchelor2000}%
\begin{APACrefauthors}%
Batchelor, G\BPBI K.%
\end{APACrefauthors}%
\unskip\
\newblock
\APACrefYear{2000}.
\newblock
\APACrefbtitle {{An Introduction to Fluid Dynamics}} {{An Introduction to Fluid
  Dynamics}}.
\newblock
\APACaddressPublisher{}{Cambridge University Press}.
\newblock
\begin{APACrefDOI} \doi{10.1017/CBO9780511800955} \end{APACrefDOI}
\PrintBackRefs{\CurrentBib}

\bibitem [\protect \citeauthoryear {%
Brandon%
, Graham%
, Waight%
\BCBL {}\ \BBA {} Gautason%
}{%
Brandon%
\ \protect \BOthers {.}}{%
{\protect \APACyear {2007}}%
}]{%
Brandon2007}
\APACinsertmetastar {%
Brandon2007}%
\begin{APACrefauthors}%
Brandon, A\BPBI D.%
, Graham, D\BPBI W.%
, Waight, T.%
\BCBL {}\ \BBA {} Gautason, B.%
\end{APACrefauthors}%
\unskip\
\newblock
\APACrefYearMonthDay{2007}{}{}.
\newblock
{\BBOQ}\APACrefatitle {{{$^{186}$}Os and {$^{187}$}Os enrichments and
  high-{$^{3}$}He/{$^{4}$}He sources in the Earth's mantle: Evidence from
  Icelandic picrites}} {{{$^{186}$}Os and {$^{187}$}Os enrichments and
  high-{$^{3}$}He/{$^{4}$}He sources in the Earth's mantle: Evidence from
  Icelandic picrites}}.{\BBCQ}
\newblock
\APACjournalVolNumPages{Geochimica et Cosmochimica Acta}{71}{18}{4570--4591}.
\newblock
\begin{APACrefDOI} \doi{10.1016/j.gca.2007.07.015} \end{APACrefDOI}
\PrintBackRefs{\CurrentBib}

\bibitem [\protect \citeauthoryear {%
Brandon%
, Snow%
, Walker%
, Morgan%
\BCBL {}\ \BBA {} Mock%
}{%
Brandon%
\ \protect \BOthers {.}}{%
{\protect \APACyear {2000}}%
}]{%
Brandon2000}
\APACinsertmetastar {%
Brandon2000}%
\begin{APACrefauthors}%
Brandon, A\BPBI D.%
, Snow, J\BPBI E.%
, Walker, R\BPBI J.%
, Morgan, J\BPBI W.%
\BCBL {}\ \BBA {} Mock, T\BPBI D.%
\end{APACrefauthors}%
\unskip\
\newblock
\APACrefYearMonthDay{2000}{}{}.
\newblock
{\BBOQ}\APACrefatitle {{{$^{190}$}Pt-{$^{186}$}Os and {$^{187}$}Re-{$^{187}$}Os
  systematics of abyssal peridotites}} {{{$^{190}$}Pt-{$^{186}$}Os and
  {$^{187}$}Re-{$^{187}$}Os systematics of abyssal peridotites}}.{\BBCQ}
\newblock
\APACjournalVolNumPages{Earth and Planetary Science Letters}{177}{3}{319--335}.
\newblock
\begin{APACrefDOI} \doi{10.1016/S0012-821X(00)00044-3} \end{APACrefDOI}
\PrintBackRefs{\CurrentBib}

\bibitem [\protect \citeauthoryear {%
Breddam%
, Kurz%
\BCBL {}\ \BBA {} Storey%
}{%
Breddam%
\ \protect \BOthers {.}}{%
{\protect \APACyear {2000}}%
}]{%
Breddam2000}
\APACinsertmetastar {%
Breddam2000}%
\begin{APACrefauthors}%
Breddam, K.%
, Kurz, M\BPBI D.%
\BCBL {}\ \BBA {} Storey, M.%
\end{APACrefauthors}%
\unskip\
\newblock
\APACrefYearMonthDay{2000}{}{}.
\newblock
{\BBOQ}\APACrefatitle {{Mapping out the conduit of the Iceland mantle plume
  with helium isotopes}} {{Mapping out the conduit of the Iceland mantle plume
  with helium isotopes}}.{\BBCQ}
\newblock
\APACjournalVolNumPages{Earth and Planetary Science Letters}{176}{1}{45--55}.
\newblock
\begin{APACrefDOI} \doi{10.1016/S0012-821X(99)00313-1} \end{APACrefDOI}
\PrintBackRefs{\CurrentBib}

\bibitem [\protect \citeauthoryear {%
Burley%
\ \BBA {} Katz%
}{%
Burley%
\ \BBA {} Katz%
}{%
{\protect \APACyear {2015}}%
}]{%
Burley2015}
\APACinsertmetastar {%
Burley2015}%
\begin{APACrefauthors}%
Burley, J\BPBI M.%
\BCBT {}\ \BBA {} Katz, R\BPBI F.%
\end{APACrefauthors}%
\unskip\
\newblock
\APACrefYearMonthDay{2015}{}{}.
\newblock
{\BBOQ}\APACrefatitle {{Variations in mid-ocean ridge CO2 emissions driven by
  glacial cycles}} {{Variations in mid-ocean ridge CO2 emissions driven by
  glacial cycles}}.{\BBCQ}
\newblock
\APACjournalVolNumPages{Earth and Planetary Science Letters}{426}{}{246--258}.
\newblock
\begin{APACrefDOI} \doi{10.1016/j.epsl.2015.06.031} \end{APACrefDOI}
\PrintBackRefs{\CurrentBib}

\bibitem [\protect \citeauthoryear {%
Burnard%
\ \BBA {} Harrison%
}{%
Burnard%
\ \BBA {} Harrison%
}{%
{\protect \APACyear {2005}}%
}]{%
Burnard2005}
\APACinsertmetastar {%
Burnard2005}%
\begin{APACrefauthors}%
Burnard, P.%
\BCBT {}\ \BBA {} Harrison, D.%
\end{APACrefauthors}%
\unskip\
\newblock
\APACrefYearMonthDay{2005}{}{}.
\newblock
{\BBOQ}\APACrefatitle {{Argon isotope constraints on modification of oxygen
  isotopes in Iceland Basalts by surficial processes}} {{Argon isotope
  constraints on modification of oxygen isotopes in Iceland Basalts by
  surficial processes}}.{\BBCQ}
\newblock
\APACjournalVolNumPages{Chemical Geology}{216}{1}{143--156}.
\newblock
\begin{APACrefDOI} \doi{10.1016/j.chemgeo.2004.11.001} \end{APACrefDOI}
\PrintBackRefs{\CurrentBib}

\bibitem [\protect \citeauthoryear {%
Chauvel%
\ \BBA {} H{\'{e}}mond%
}{%
Chauvel%
\ \BBA {} H{\'{e}}mond%
}{%
{\protect \APACyear {2000}}%
}]{%
Chauvel2000}
\APACinsertmetastar {%
Chauvel2000}%
\begin{APACrefauthors}%
Chauvel, C.%
\BCBT {}\ \BBA {} H{\'{e}}mond, C.%
\end{APACrefauthors}%
\unskip\
\newblock
\APACrefYearMonthDay{2000}{}{}.
\newblock
{\BBOQ}\APACrefatitle {{Melting of a complete section of recycled oceanic
  crust: Trace element and Pb isotopic evidence from Iceland}} {{Melting of a
  complete section of recycled oceanic crust: Trace element and Pb isotopic
  evidence from Iceland}}.{\BBCQ}
\newblock
\APACjournalVolNumPages{Geochemistry, Geophysics, Geosystems}{1}{2}{1--22}.
\newblock
\begin{APACrefDOI} \doi{10.1029/1999GC000002} \end{APACrefDOI}
\PrintBackRefs{\CurrentBib}

\bibitem [\protect \citeauthoryear {%
Condomines%
\ \protect \BOthers {.}}{%
Condomines%
\ \protect \BOthers {.}}{%
{\protect \APACyear {1983}}%
}]{%
Condomines1983}
\APACinsertmetastar {%
Condomines1983}%
\begin{APACrefauthors}%
Condomines, M.%
, Grönvold, K.%
, Hooker, P.%
, Muehlenbachs, K.%
, O'Nions, R.%
, Óskarsson, N.%
\BCBL {}\ \BBA {} Oxburgh, E.%
\end{APACrefauthors}%
\unskip\
\newblock
\APACrefYearMonthDay{1983}{}{}.
\newblock
{\BBOQ}\APACrefatitle {{Helium, oxygen, strontium and neodymium isotopic
  relationships in Icelandic volcanics}} {{Helium, oxygen, strontium and
  neodymium isotopic relationships in Icelandic volcanics}}.{\BBCQ}
\newblock
\APACjournalVolNumPages{Earth and Planetary Science Letters}{66}{}{125--136}.
\newblock
\begin{APACrefDOI} \doi{10.1016/0012-821X(83)90131-0} \end{APACrefDOI}
\PrintBackRefs{\CurrentBib}

\bibitem [\protect \citeauthoryear {%
Crowley%
, Katz%
, Huybers%
, Langmuir%
\BCBL {}\ \BBA {} Park%
}{%
Crowley%
\ \protect \BOthers {.}}{%
{\protect \APACyear {2015}}%
}]{%
Crowley2015}
\APACinsertmetastar {%
Crowley2015}%
\begin{APACrefauthors}%
Crowley, J\BPBI W.%
, Katz, R\BPBI F.%
, Huybers, P.%
, Langmuir, C\BPBI H.%
\BCBL {}\ \BBA {} Park, S\BHBI H.%
\end{APACrefauthors}%
\unskip\
\newblock
\APACrefYearMonthDay{2015}{}{}.
\newblock
{\BBOQ}\APACrefatitle {{Glacial cycles drive variations in the production of
  oceanic crust}} {{Glacial cycles drive variations in the production of
  oceanic crust}}.{\BBCQ}
\newblock
\APACjournalVolNumPages{Science}{347}{6227}{1237--1240}.
\newblock
\begin{APACrefDOI} \doi{10.1126/science.1261508} \end{APACrefDOI}
\PrintBackRefs{\CurrentBib}

\bibitem [\protect \citeauthoryear {%
Darbyshire%
\ \protect \BOthers {.}}{%
Darbyshire%
\ \protect \BOthers {.}}{%
{\protect \APACyear {2000}}%
}]{%
Darbyshire2000}
\APACinsertmetastar {%
Darbyshire2000}%
\begin{APACrefauthors}%
Darbyshire, F\BPBI A.%
, Priestley, K\BPBI F.%
, White, R\BPBI S.%
, Stefánsson, R.%
, Gudmundsson, G\BPBI B.%
\BCBL {}\ \BBA {} Jakobsdóttir, S\BPBI S.%
\end{APACrefauthors}%
\unskip\
\newblock
\APACrefYearMonthDay{2000}{}{}.
\newblock
{\BBOQ}\APACrefatitle {{Crustal structure of central and northern Iceland from
  analysis of teleseismic receiver functions}} {{Crustal structure of central
  and northern Iceland from analysis of teleseismic receiver
  functions}}.{\BBCQ}
\newblock
\APACjournalVolNumPages{Geophysical Journal International}{143}{1}{163--184}.
\newblock
\begin{APACrefDOI} \doi{10.1046/j.1365-246x.2000.00224.x} \end{APACrefDOI}
\PrintBackRefs{\CurrentBib}

\bibitem [\protect \citeauthoryear {%
Debaille%
\ \protect \BOthers {.}}{%
Debaille%
\ \protect \BOthers {.}}{%
{\protect \APACyear {2009}}%
}]{%
Debaille2009}
\APACinsertmetastar {%
Debaille2009}%
\begin{APACrefauthors}%
Debaille, V.%
, Tr{\o}nnes, R\BPBI G.%
, Brandon, A\BPBI D.%
, Waight, T\BPBI E.%
, Graham, D\BPBI W.%
\BCBL {}\ \BBA {} Lee, C\BPBI T\BPBI A.%
\end{APACrefauthors}%
\unskip\
\newblock
\APACrefYearMonthDay{2009}{}{}.
\newblock
{\BBOQ}\APACrefatitle {{Primitive off-rift basalts from Iceland and Jan Mayen:
  Os-isotopic evidence for a mantle source containing enriched subcontinental
  lithosphere}} {{Primitive off-rift basalts from Iceland and Jan Mayen:
  Os-isotopic evidence for a mantle source containing enriched subcontinental
  lithosphere}}.{\BBCQ}
\newblock
\APACjournalVolNumPages{Geochimica et Cosmochimica Acta}{73}{11}{3423--3449}.
\newblock
\begin{APACrefDOI} \doi{10.1016/j.gca.2009.03.002} \end{APACrefDOI}
\PrintBackRefs{\CurrentBib}

\bibitem [\protect \citeauthoryear {%
Dixon%
}{%
Dixon%
}{%
{\protect \APACyear {2003}}%
}]{%
Dixon2003}
\APACinsertmetastar {%
Dixon2003}%
\begin{APACrefauthors}%
Dixon, E\BPBI T.%
\end{APACrefauthors}%
\unskip\
\newblock
\APACrefYearMonthDay{2003}{}{}.
\newblock
{\BBOQ}\APACrefatitle {{Interpretation of helium and neon isotopic
  heterogeneity in Icelandic basalts}} {{Interpretation of helium and neon
  isotopic heterogeneity in Icelandic basalts}}.{\BBCQ}
\newblock
\APACjournalVolNumPages{Earth and Planetary Science Letters}{206}{1-2}{83--99}.
\newblock
\begin{APACrefDOI} \doi{10.1016/S0012-821X(02)01071-3} \end{APACrefDOI}
\PrintBackRefs{\CurrentBib}

\bibitem [\protect \citeauthoryear {%
Dixon%
, Honda%
, McDougall%
, Campbell%
\BCBL {}\ \BBA {} Sigurdsson%
}{%
Dixon%
\ \protect \BOthers {.}}{%
{\protect \APACyear {2000}}%
}]{%
Dixon2000}
\APACinsertmetastar {%
Dixon2000}%
\begin{APACrefauthors}%
Dixon, E\BPBI T.%
, Honda, M.%
, McDougall, I.%
, Campbell, I\BPBI H.%
\BCBL {}\ \BBA {} Sigurdsson, I.%
\end{APACrefauthors}%
\unskip\
\newblock
\APACrefYearMonthDay{2000}{}{}.
\newblock
{\BBOQ}\APACrefatitle {{Preservation of near-solar neon isotopic ratios in
  Icelandic basalts}} {{Preservation of near-solar neon isotopic ratios in
  Icelandic basalts}}.{\BBCQ}
\newblock
\APACjournalVolNumPages{Earth and Planetary Science Letters}{180}{3}{309--324}.
\newblock
\begin{APACrefDOI} \doi{10.1016/S0012-821X(00)00164-3} \end{APACrefDOI}
\PrintBackRefs{\CurrentBib}

\bibitem [\protect \citeauthoryear {%
Eason%
\ \BBA {} Sinton%
}{%
Eason%
\ \BBA {} Sinton%
}{%
{\protect \APACyear {2009}}%
}]{%
Eason2009}
\APACinsertmetastar {%
Eason2009}%
\begin{APACrefauthors}%
Eason, D\BPBI E.%
\BCBT {}\ \BBA {} Sinton, J\BPBI M.%
\end{APACrefauthors}%
\unskip\
\newblock
\APACrefYearMonthDay{2009}{}{}.
\newblock
{\BBOQ}\APACrefatitle {{Lava shields and fissure eruptions of the Western
  Volcanic Zone, Iceland: Evidence for magma chambers and crustal interaction}}
  {{Lava shields and fissure eruptions of the Western Volcanic Zone, Iceland:
  Evidence for magma chambers and crustal interaction}}.{\BBCQ}
\newblock
\APACjournalVolNumPages{Journal of Volcanology and Geothermal
  Research}{186}{3}{331--348}.
\newblock
\begin{APACrefDOI} \doi{10.1016/j.jvolgeores.2009.06.009} \end{APACrefDOI}
\PrintBackRefs{\CurrentBib}

\bibitem [\protect \citeauthoryear {%
Eason%
, Sinton%
, Gr{\"{o}}nvold%
\BCBL {}\ \BBA {} Kurz%
}{%
Eason%
\ \protect \BOthers {.}}{%
{\protect \APACyear {2015}}%
}]{%
Eason2015}
\APACinsertmetastar {%
Eason2015}%
\begin{APACrefauthors}%
Eason, D\BPBI E.%
, Sinton, J\BPBI M.%
, Gr{\"{o}}nvold, K.%
\BCBL {}\ \BBA {} Kurz, M\BPBI D.%
\end{APACrefauthors}%
\unskip\
\newblock
\APACrefYearMonthDay{2015}{}{}.
\newblock
{\BBOQ}\APACrefatitle {{Effects of deglaciation on the petrology and eruptive
  history of the Western Volcanic Zone, Iceland}} {{Effects of deglaciation on
  the petrology and eruptive history of the Western Volcanic Zone,
  Iceland}}.{\BBCQ}
\newblock
\APACjournalVolNumPages{Bulletin of Volcanology}{77}{6}{}.
\newblock
\begin{APACrefDOI} \doi{10.1007/s00445-015-0916-0} \end{APACrefDOI}
\PrintBackRefs{\CurrentBib}

\bibitem [\protect \citeauthoryear {%
Elliott%
, Hawkesworth%
\BCBL {}\ \BBA {} Gr{\"{o}}nvold%
}{%
Elliott%
\ \protect \BOthers {.}}{%
{\protect \APACyear {1991}}%
}]{%
Elliott1991}
\APACinsertmetastar {%
Elliott1991}%
\begin{APACrefauthors}%
Elliott, T\BPBI R.%
, Hawkesworth, C\BPBI J.%
\BCBL {}\ \BBA {} Gr{\"{o}}nvold, K.%
\end{APACrefauthors}%
\unskip\
\newblock
\APACrefYearMonthDay{1991}{}{}.
\newblock
{\BBOQ}\APACrefatitle {{Dynamic melting of the Iceland plume}} {{Dynamic
  melting of the Iceland plume}}.{\BBCQ}
\newblock
\APACjournalVolNumPages{Nature}{351}{6323}{201--206}.
\newblock
\begin{APACrefDOI} \doi{10.1038/351201a0} \end{APACrefDOI}
\PrintBackRefs{\CurrentBib}

\bibitem [\protect \citeauthoryear {%
Fitton%
, Saunders%
, Kempton%
\BCBL {}\ \BBA {} Hardarson%
}{%
Fitton%
\ \protect \BOthers {.}}{%
{\protect \APACyear {2003}}%
}]{%
Fitton2003}
\APACinsertmetastar {%
Fitton2003}%
\begin{APACrefauthors}%
Fitton, J\BPBI G.%
, Saunders, A\BPBI D.%
, Kempton, P\BPBI D.%
\BCBL {}\ \BBA {} Hardarson, B\BPBI S.%
\end{APACrefauthors}%
\unskip\
\newblock
\APACrefYearMonthDay{2003}{}{}.
\newblock
{\BBOQ}\APACrefatitle {{Does depleted mantle form an intrinsic part of the
  Iceland plume?}} {{Does depleted mantle form an intrinsic part of the Iceland
  plume?}}{\BBCQ}
\newblock
\APACjournalVolNumPages{Geochemistry, Geophysics, Geosystems}{4}{3}{1--14}.
\newblock
\begin{APACrefDOI} \doi{10.1029/2002GC000424} \end{APACrefDOI}
\PrintBackRefs{\CurrentBib}

\bibitem [\protect \citeauthoryear {%
F{\"u}ri%
\ \protect \BOthers {.}}{%
F{\"u}ri%
\ \protect \BOthers {.}}{%
{\protect \APACyear {2010}}%
}]{%
Furi2010}
\APACinsertmetastar {%
Furi2010}%
\begin{APACrefauthors}%
F{\"u}ri, E.%
, Hilton, D.%
, Halld{\'o}rsson, S.%
, Barry, P.%
, Hahm, D.%
, Fischer, T.%
\BCBL {}\ \BBA {} Gr{\"o}nvold, K.%
\end{APACrefauthors}%
\unskip\
\newblock
\APACrefYearMonthDay{2010}{}{}.
\newblock
{\BBOQ}\APACrefatitle {{Apparent decoupling of the He and Ne isotope
  systematics of the Icelandic mantle: The role of He depletion, melt mixing,
  degassing fractionation and air interaction}} {{Apparent decoupling of the He
  and Ne isotope systematics of the Icelandic mantle: The role of He depletion,
  melt mixing, degassing fractionation and air interaction}}.{\BBCQ}
\newblock
\APACjournalVolNumPages{Geochimica et Cosmochimica Acta}{74}{11}{3307--3332}.
\newblock
\begin{APACrefDOI} \doi{10.1016/j.gca.2010.03.023} \end{APACrefDOI}
\PrintBackRefs{\CurrentBib}

\bibitem [\protect \citeauthoryear {%
Gee%
, Taylor%
, Thirlwall%
\BCBL {}\ \BBA {} Murton%
}{%
Gee%
, Taylor%
\BCBL {}\ \protect \BOthers {.}}{%
{\protect \APACyear {1998}}%
}]{%
Gee1998a}
\APACinsertmetastar {%
Gee1998a}%
\begin{APACrefauthors}%
Gee, M\BPBI A\BPBI M.%
, Taylor, R\BPBI N.%
, Thirlwall, M\BPBI F.%
\BCBL {}\ \BBA {} Murton, B\BPBI J.%
\end{APACrefauthors}%
\unskip\
\newblock
\APACrefYearMonthDay{1998}{}{}.
\newblock
{\BBOQ}\APACrefatitle {{Glacioisostacy controls chemical and isotopic
  characteristics of tholeiites from the Reykjanes Peninsula, SW Iceland}}
  {{Glacioisostacy controls chemical and isotopic characteristics of tholeiites
  from the Reykjanes Peninsula, SW Iceland}}.{\BBCQ}
\newblock
\APACjournalVolNumPages{Earth and Planetary Science Letters}{164}{1-2}{1--5}.
\newblock
\begin{APACrefDOI} \doi{10.1016/S0012-821X(98)00246-5} \end{APACrefDOI}
\PrintBackRefs{\CurrentBib}

\bibitem [\protect \citeauthoryear {%
Gee%
, Thirlwall%
, Taylor%
, Lowry%
\BCBL {}\ \BBA {} Murton%
}{%
Gee%
, Thirlwall%
\BCBL {}\ \protect \BOthers {.}}{%
{\protect \APACyear {1998}}%
}]{%
Gee1998b}
\APACinsertmetastar {%
Gee1998b}%
\begin{APACrefauthors}%
Gee, M\BPBI A\BPBI M.%
, Thirlwall, M\BPBI F.%
, Taylor, R\BPBI N.%
, Lowry, D.%
\BCBL {}\ \BBA {} Murton, B\BPBI J.%
\end{APACrefauthors}%
\unskip\
\newblock
\APACrefYearMonthDay{1998}{}{}.
\newblock
{\BBOQ}\APACrefatitle {{Crustal Processes: Major Controls on Reykjanes
  Peninsula Lava Chemistry, SW Iceland}} {{Crustal Processes: Major Controls on
  Reykjanes Peninsula Lava Chemistry, SW Iceland}}.{\BBCQ}
\newblock
\APACjournalVolNumPages{Journal of Petrology}{39}{5}{819--839}.
\newblock
\begin{APACrefDOI} \doi{10.1093/petroj/39.5.819} \end{APACrefDOI}
\PrintBackRefs{\CurrentBib}

\bibitem [\protect \citeauthoryear {%
Geirsd{\'o}ttir%
, Miller%
, Axford%
\BCBL {}\ \BBA {} {\'O}lafsd{\'o}ttir%
}{%
Geirsd{\'o}ttir%
\ \protect \BOthers {.}}{%
{\protect \APACyear {2009}}%
}]{%
Geirsdottir2009}
\APACinsertmetastar {%
Geirsdottir2009}%
\begin{APACrefauthors}%
Geirsd{\'o}ttir, {\'A}.%
, Miller, G\BPBI H.%
, Axford, Y.%
\BCBL {}\ \BBA {} {\'O}lafsd{\'o}ttir, S.%
\end{APACrefauthors}%
\unskip\
\newblock
\APACrefYearMonthDay{2009}{}{}.
\newblock
{\BBOQ}\APACrefatitle {{Holocene and latest Pleistocene climate and glacier
  fluctuations in Iceland}} {{Holocene and latest Pleistocene climate and
  glacier fluctuations in Iceland}}.{\BBCQ}
\newblock
\APACjournalVolNumPages{Quaternary Science Reviews}{28}{21}{2107--2118}.
\newblock
\APACrefnote{{Holocene and Latest Pleistocene Alpine Glacier Fluctuations: A
  Global Perspective}}
\newblock
\begin{APACrefDOI} \doi{10.1016/j.quascirev.2009.03.013} \end{APACrefDOI}
\PrintBackRefs{\CurrentBib}

\bibitem [\protect \citeauthoryear {%
Ghiorso%
, Hirschmann%
, Reiners%
\BCBL {}\ \BBA {} Kress~III%
}{%
Ghiorso%
\ \protect \BOthers {.}}{%
{\protect \APACyear {2002}}%
}]{%
Ghiorso2002}
\APACinsertmetastar {%
Ghiorso2002}%
\begin{APACrefauthors}%
Ghiorso, M\BPBI S.%
, Hirschmann, M\BPBI M.%
, Reiners, P\BPBI W.%
\BCBL {}\ \BBA {} Kress~III, V\BPBI C.%
\end{APACrefauthors}%
\unskip\
\newblock
\APACrefYearMonthDay{2002}{}{}.
\newblock
{\BBOQ}\APACrefatitle {{The pMELTS: A revision of MELTS for improved
  calculation of phase relations and major element partitioning related to
  partial melting of the mantle to 3 GPa}} {{The pMELTS: A revision of MELTS
  for improved calculation of phase relations and major element partitioning
  related to partial melting of the mantle to 3 GPa}}.{\BBCQ}
\newblock
\APACjournalVolNumPages{Geochemistry, Geophysics, Geosystems}{3}{5}{1--35}.
\newblock
\begin{APACrefDOI} \doi{10.1029/2001GC000217} \end{APACrefDOI}
\PrintBackRefs{\CurrentBib}

\bibitem [\protect \citeauthoryear {%
Hardarson%
, Fitton%
, Ellam%
\BCBL {}\ \BBA {} Pringle%
}{%
Hardarson%
\ \protect \BOthers {.}}{%
{\protect \APACyear {1997}}%
}]{%
Hardarson1997}
\APACinsertmetastar {%
Hardarson1997}%
\begin{APACrefauthors}%
Hardarson, B.%
, Fitton, J.%
, Ellam, R.%
\BCBL {}\ \BBA {} Pringle, M.%
\end{APACrefauthors}%
\unskip\
\newblock
\APACrefYearMonthDay{1997}{}{}.
\newblock
{\BBOQ}\APACrefatitle {{Rift relocation — A geochemical and geochronological
  investigation of a palaeo-rift in northwest Iceland}} {{Rift relocation — A
  geochemical and geochronological investigation of a palaeo-rift in northwest
  Iceland}}.{\BBCQ}
\newblock
\APACjournalVolNumPages{Earth and Planetary Science
  Letters}{153}{3-4}{181--196}.
\newblock
\begin{APACrefDOI} \doi{10.1016/S0012-821X(97)00145-3} \end{APACrefDOI}
\PrintBackRefs{\CurrentBib}

\bibitem [\protect \citeauthoryear {%
Hemond%
\ \protect \BOthers {.}}{%
Hemond%
\ \protect \BOthers {.}}{%
{\protect \APACyear {1993}}%
}]{%
Hemond1993}
\APACinsertmetastar {%
Hemond1993}%
\begin{APACrefauthors}%
Hemond, C.%
, Arndt, N\BPBI T.%
, Lichtenstein, U.%
, Hofmann, A\BPBI W.%
, Oskarsson, N.%
\BCBL {}\ \BBA {} Steinthorsson, S.%
\end{APACrefauthors}%
\unskip\
\newblock
\APACrefYearMonthDay{1993}{}{}.
\newblock
{\BBOQ}\APACrefatitle {{The heterogeneous Iceland plume: Nd-Sr-O isotopes and
  trace element constraints}} {{The heterogeneous Iceland plume: Nd-Sr-O
  isotopes and trace element constraints}}.{\BBCQ}
\newblock
\APACjournalVolNumPages{Journal of Geophysical Research}{98}{B9}{15833}.
\newblock
\begin{APACrefDOI} \doi{10.1029/93JB01093} \end{APACrefDOI}
\PrintBackRefs{\CurrentBib}

\bibitem [\protect \citeauthoryear {%
Hubbard%
, Sugden%
, Dugmore%
, Norddahl%
\BCBL {}\ \BBA {} P{\'{e}}tursson%
}{%
Hubbard%
\ \protect \BOthers {.}}{%
{\protect \APACyear {2006}}%
}]{%
Hubbard2006}
\APACinsertmetastar {%
Hubbard2006}%
\begin{APACrefauthors}%
Hubbard, A.%
, Sugden, D.%
, Dugmore, A.%
, Norddahl, H.%
\BCBL {}\ \BBA {} P{\'{e}}tursson, H\BPBI G.%
\end{APACrefauthors}%
\unskip\
\newblock
\APACrefYearMonthDay{2006}{}{}.
\newblock
{\BBOQ}\APACrefatitle {{A modelling insight into the Icelandic Last Glacial
  Maximum ice sheet}} {{A modelling insight into the Icelandic Last Glacial
  Maximum ice sheet}}.{\BBCQ}
\newblock
\APACjournalVolNumPages{Quaternary Science Reviews}{25}{17-18}{2283--2296}.
\newblock
\begin{APACrefDOI} \doi{10.1016/j.quascirev.2006.04.001} \end{APACrefDOI}
\PrintBackRefs{\CurrentBib}

\bibitem [\protect \citeauthoryear {%
Huppert%
}{%
Huppert%
}{%
{\protect \APACyear {1982}}%
}]{%
Huppert1982}
\APACinsertmetastar {%
Huppert1982}%
\begin{APACrefauthors}%
Huppert, H\BPBI E.%
\end{APACrefauthors}%
\unskip\
\newblock
\APACrefYearMonthDay{1982}{}{}.
\newblock
{\BBOQ}\APACrefatitle {{The propagation of two-dimensional and axisymmetric
  viscous gravity currents over a rigid horizontal surface}} {{The propagation
  of two-dimensional and axisymmetric viscous gravity currents over a rigid
  horizontal surface}}.{\BBCQ}
\newblock
\APACjournalVolNumPages{Journal of Fluid Mechanics}{121}{}{43--58}.
\newblock
\begin{APACrefDOI} \doi{10.1017/S0022112082001797} \end{APACrefDOI}
\PrintBackRefs{\CurrentBib}

\bibitem [\protect \citeauthoryear {%
Jakobsson%
, J{\'o}nsson%
\BCBL {}\ \BBA {} Shido%
}{%
Jakobsson%
\ \protect \BOthers {.}}{%
{\protect \APACyear {1978}}%
}]{%
Jakobsson1978}
\APACinsertmetastar {%
Jakobsson1978}%
\begin{APACrefauthors}%
Jakobsson, S\BPBI P.%
, J{\'o}nsson, J.%
\BCBL {}\ \BBA {} Shido, F.%
\end{APACrefauthors}%
\unskip\
\newblock
\APACrefYearMonthDay{1978}{}{}.
\newblock
{\BBOQ}\APACrefatitle {{Petrology of the Western Reykjanes Peninsula, Iceland}}
  {{Petrology of the Western Reykjanes Peninsula, Iceland}}.{\BBCQ}
\newblock
\APACjournalVolNumPages{Journal of Petrology}{19}{4}{669--705}.
\newblock
\begin{APACrefDOI} \doi{10.1093/petrology/19.4.669} \end{APACrefDOI}
\PrintBackRefs{\CurrentBib}

\bibitem [\protect \citeauthoryear {%
J{\'{o}}nasson%
}{%
J{\'{o}}nasson%
}{%
{\protect \APACyear {2005}}%
}]{%
Jonasson2005}
\APACinsertmetastar {%
Jonasson2005}%
\begin{APACrefauthors}%
J{\'{o}}nasson, K.%
\end{APACrefauthors}%
\unskip\
\newblock
\APACrefYearMonthDay{2005}{}{}.
\newblock
{\BBOQ}\APACrefatitle {{Magmatic evolution of the Hei{\dh}arspor{\dh}ur ridge,
  NE-Iceland}} {{Magmatic evolution of the Hei{\dh}arspor{\dh}ur ridge,
  NE-Iceland}}.{\BBCQ}
\newblock
\APACjournalVolNumPages{Journal of Volcanology and Geothermal
  Research}{147}{1-2}{109--124}.
\newblock
\begin{APACrefDOI} \doi{10.1016/j.jvolgeores.2005.03.009} \end{APACrefDOI}
\PrintBackRefs{\CurrentBib}

\bibitem [\protect \citeauthoryear {%
Jull%
\ \BBA {} McKenzie%
}{%
Jull%
\ \BBA {} McKenzie%
}{%
{\protect \APACyear {1996}}%
}]{%
Jull1996}
\APACinsertmetastar {%
Jull1996}%
\begin{APACrefauthors}%
Jull, M.%
\BCBT {}\ \BBA {} McKenzie, D.%
\end{APACrefauthors}%
\unskip\
\newblock
\APACrefYearMonthDay{1996}{}{}.
\newblock
{\BBOQ}\APACrefatitle {{The effect of deglaciation on mantle melting beneath
  Iceland}} {{The effect of deglaciation on mantle melting beneath
  Iceland}}.{\BBCQ}
\newblock
\APACjournalVolNumPages{Journal of Geophysical Research: Solid
  Earth}{101}{B10}{21815--21828}.
\newblock
\begin{APACrefDOI} \doi{10.1029/96JB01308} \end{APACrefDOI}
\PrintBackRefs{\CurrentBib}

\bibitem [\protect \citeauthoryear {%
Katz%
, Spiegelman%
\BCBL {}\ \BBA {} Langmuir%
}{%
Katz%
\ \protect \BOthers {.}}{%
{\protect \APACyear {2003}}%
}]{%
Katz2003}
\APACinsertmetastar {%
Katz2003}%
\begin{APACrefauthors}%
Katz, R\BPBI F.%
, Spiegelman, M.%
\BCBL {}\ \BBA {} Langmuir, C\BPBI H.%
\end{APACrefauthors}%
\unskip\
\newblock
\APACrefYearMonthDay{2003}{}{}.
\newblock
{\BBOQ}\APACrefatitle {{A new parameterization of hydrous mantle melting}} {{A
  new parameterization of hydrous mantle melting}}.{\BBCQ}
\newblock
\APACjournalVolNumPages{Geochemistry, Geophysics, Geosystems}{4}{9}{}.
\newblock
\begin{APACrefDOI} \doi{10.1029/2002GC000433} \end{APACrefDOI}
\PrintBackRefs{\CurrentBib}

\bibitem [\protect \citeauthoryear {%
Kelemen%
, Hirth%
, Shimizu%
, Spiegelman%
\BCBL {}\ \BBA {} Dick%
}{%
Kelemen%
\ \protect \BOthers {.}}{%
{\protect \APACyear {1997}}%
}]{%
Kelemen1997}
\APACinsertmetastar {%
Kelemen1997}%
\begin{APACrefauthors}%
Kelemen, P\BPBI B.%
, Hirth, G.%
, Shimizu, N.%
, Spiegelman, M.%
\BCBL {}\ \BBA {} Dick, H\BPBI J\BPBI B.%
\end{APACrefauthors}%
\unskip\
\newblock
\APACrefYearMonthDay{1997}{}{}.
\newblock
{\BBOQ}\APACrefatitle {{A review of melt migration processes in the
  adiabatically upwelling mantle beneath oceanic spreading ridges}} {{A review
  of melt migration processes in the adiabatically upwelling mantle beneath
  oceanic spreading ridges}}.{\BBCQ}
\newblock
\APACjournalVolNumPages{Philosophical Transactions of the Royal Society of
  London. Series A: Mathematical, Physical and Engineering
  Sciences}{355}{1723}{283--318}.
\newblock
\begin{APACrefDOI} \doi{10.1098/rsta.1997.0010} \end{APACrefDOI}
\PrintBackRefs{\CurrentBib}

\bibitem [\protect \citeauthoryear {%
Kempton%
\ \protect \BOthers {.}}{%
Kempton%
\ \protect \BOthers {.}}{%
{\protect \APACyear {2000}}%
}]{%
Kempton2000}
\APACinsertmetastar {%
Kempton2000}%
\begin{APACrefauthors}%
Kempton, P.%
, Fitton, J.%
, Saunders, A.%
, Nowell, G.%
, Taylor, R.%
, Hardarson, B.%
\BCBL {}\ \BBA {} Pearson, G.%
\end{APACrefauthors}%
\unskip\
\newblock
\APACrefYearMonthDay{2000}{}{}.
\newblock
{\BBOQ}\APACrefatitle {{The Iceland plume in space and time: a Sr-Nd-Pb-Hf
  study of the North Atlantic rifted margin}} {{The Iceland plume in space and
  time: a Sr-Nd-Pb-Hf study of the North Atlantic rifted margin}}.{\BBCQ}
\newblock
\APACjournalVolNumPages{Earth and Planetary Science Letters}{177}{3}{255--271}.
\newblock
\begin{APACrefDOI} \doi{10.1016/S0012-821X(00)00047-9} \end{APACrefDOI}
\PrintBackRefs{\CurrentBib}

\bibitem [\protect \citeauthoryear {%
Kokfelt%
\ \protect \BOthers {.}}{%
Kokfelt%
\ \protect \BOthers {.}}{%
{\protect \APACyear {2006}}%
}]{%
Kokfelt2006}
\APACinsertmetastar {%
Kokfelt2006}%
\begin{APACrefauthors}%
Kokfelt, T\BPBI F.%
, Hoernle, K.%
, Hauff, F.%
, Fiebig, J.%
, Werner, R.%
\BCBL {}\ \BBA {} Garbe-Sch{\"{o}}nberg, D.%
\end{APACrefauthors}%
\unskip\
\newblock
\APACrefYearMonthDay{2006}{}{}.
\newblock
{\BBOQ}\APACrefatitle {{Combined trace element and Pb-Nd-Sr-O isotope evidence
  for recycled oceanic crust (upper and lower) in the Iceland mantle plume}}
  {{Combined trace element and Pb-Nd-Sr-O isotope evidence for recycled oceanic
  crust (upper and lower) in the Iceland mantle plume}}.{\BBCQ}
\newblock
\APACjournalVolNumPages{Journal of Petrology}{47}{9}{1705--1749}.
\newblock
\begin{APACrefDOI} \doi{10.1093/petrology/egl025} \end{APACrefDOI}
\PrintBackRefs{\CurrentBib}

\bibitem [\protect \citeauthoryear {%
Koornneef%
, Stracke%
, Bourdon%
\BCBL {}\ \BBA {} Gr{\"{o}}nvold%
}{%
Koornneef%
\ \protect \BOthers {.}}{%
{\protect \APACyear {2012}}%
}]{%
Koornneef2012}
\APACinsertmetastar {%
Koornneef2012}%
\begin{APACrefauthors}%
Koornneef, J\BPBI M.%
, Stracke, A.%
, Bourdon, B.%
\BCBL {}\ \BBA {} Gr{\"{o}}nvold, K.%
\end{APACrefauthors}%
\unskip\
\newblock
\APACrefYearMonthDay{2012}{}{}.
\newblock
{\BBOQ}\APACrefatitle {{The influence of source heterogeneity on the U-Th-Pa-Ra
  disequilibria in post-glacial tholeiites from Iceland}} {{The influence of
  source heterogeneity on the U-Th-Pa-Ra disequilibria in post-glacial
  tholeiites from Iceland}}.{\BBCQ}
\newblock
\APACjournalVolNumPages{Geochimica et Cosmochimica Acta}{87}{}{243--266}.
\newblock
\begin{APACrefDOI} \doi{10.1016/j.gca.2012.03.041} \end{APACrefDOI}
\PrintBackRefs{\CurrentBib}

\bibitem [\protect \citeauthoryear {%
Kurz%
, Meyer%
\BCBL {}\ \BBA {} Sigurdsson%
}{%
Kurz%
\ \protect \BOthers {.}}{%
{\protect \APACyear {1985}}%
}]{%
Kurz1985}
\APACinsertmetastar {%
Kurz1985}%
\begin{APACrefauthors}%
Kurz, M\BPBI D.%
, Meyer, P\BPBI S.%
\BCBL {}\ \BBA {} Sigurdsson, H.%
\end{APACrefauthors}%
\unskip\
\newblock
\APACrefYearMonthDay{1985}{}{}.
\newblock
{\BBOQ}\APACrefatitle {{Helium isotopic systematics within the neovolcanic
  zones of Iceland}} {{Helium isotopic systematics within the neovolcanic zones
  of Iceland}}.{\BBCQ}
\newblock
\APACjournalVolNumPages{Earth and Planetary Science Letters}{74}{4}{291--305}.
\newblock
\begin{APACrefDOI} \doi{10.1016/S0012-821X(85)80001-7} \end{APACrefDOI}
\PrintBackRefs{\CurrentBib}

\bibitem [\protect \citeauthoryear {%
Licciardi%
, Kurz%
\BCBL {}\ \BBA {} Curtice%
}{%
Licciardi%
\ \protect \BOthers {.}}{%
{\protect \APACyear {2007}}%
}]{%
Licciardi2007}
\APACinsertmetastar {%
Licciardi2007}%
\begin{APACrefauthors}%
Licciardi, J\BPBI M.%
, Kurz, M\BPBI D.%
\BCBL {}\ \BBA {} Curtice, J\BPBI M.%
\end{APACrefauthors}%
\unskip\
\newblock
\APACrefYearMonthDay{2007}{}{}.
\newblock
{\BBOQ}\APACrefatitle {{Glacial and volcanic history of Icelandic table
  mountains from cosmogenic {$^{3}$}He exposure ages}} {{Glacial and volcanic
  history of Icelandic table mountains from cosmogenic {$^{3}$}He exposure
  ages}}.{\BBCQ}
\newblock
\APACjournalVolNumPages{Quaternary Science Reviews}{26}{11-12}{1529--1546}.
\newblock
\begin{APACrefDOI} \doi{10.1016/j.quascirev.2007.02.016} \end{APACrefDOI}
\PrintBackRefs{\CurrentBib}

\bibitem [\protect \citeauthoryear {%
Maclennan%
}{%
Maclennan%
}{%
{\protect \APACyear {2008}}%
}]{%
Maclennan2008}
\APACinsertmetastar {%
Maclennan2008}%
\begin{APACrefauthors}%
Maclennan, J.%
\end{APACrefauthors}%
\unskip\
\newblock
\APACrefYearMonthDay{2008}{}{}.
\newblock
{\BBOQ}\APACrefatitle {{Concurrent mixing and cooling of melts under Iceland}}
  {{Concurrent mixing and cooling of melts under Iceland}}.{\BBCQ}
\newblock
\APACjournalVolNumPages{Journal of Petrology}{49}{11}{1931--1953}.
\newblock
\begin{APACrefDOI} \doi{10.1093/petrology/egn052} \end{APACrefDOI}
\PrintBackRefs{\CurrentBib}

\bibitem [\protect \citeauthoryear {%
Maclennan%
, Hulme%
\BCBL {}\ \BBA {} Singh%
}{%
Maclennan%
\ \protect \BOthers {.}}{%
{\protect \APACyear {2004}}%
}]{%
Maclennan2004}
\APACinsertmetastar {%
Maclennan2004}%
\begin{APACrefauthors}%
Maclennan, J.%
, Hulme, T.%
\BCBL {}\ \BBA {} Singh, S\BPBI C.%
\end{APACrefauthors}%
\unskip\
\newblock
\APACrefYearMonthDay{2004}{}{}.
\newblock
{\BBOQ}\APACrefatitle {{Thermal models of oceanic crustal accretion: Linking
  geophysical, geological and petrological observations}} {{Thermal models of
  oceanic crustal accretion: Linking geophysical, geological and petrological
  observations}}.{\BBCQ}
\newblock
\APACjournalVolNumPages{Geochemistry, Geophysics, Geosystems}{5}{2}{}.
\newblock
\begin{APACrefDOI} \doi{10.1029/2003GC000605} \end{APACrefDOI}
\PrintBackRefs{\CurrentBib}

\bibitem [\protect \citeauthoryear {%
Maclennan%
, Jull%
, McKenzie%
, Slater%
\BCBL {}\ \BBA {} Gr{\"{o}}nvold%
}{%
Maclennan%
\ \protect \BOthers {.}}{%
{\protect \APACyear {2002}}%
}]{%
Maclennan2002}
\APACinsertmetastar {%
Maclennan2002}%
\begin{APACrefauthors}%
Maclennan, J.%
, Jull, M.%
, McKenzie, D.%
, Slater, L.%
\BCBL {}\ \BBA {} Gr{\"{o}}nvold, K.%
\end{APACrefauthors}%
\unskip\
\newblock
\APACrefYearMonthDay{2002}{}{}.
\newblock
{\BBOQ}\APACrefatitle {{The link between volcanism and deglaciation in
  Iceland}} {{The link between volcanism and deglaciation in Iceland}}.{\BBCQ}
\newblock
\APACjournalVolNumPages{Geochemistry, Geophysics, Geosystems}{3}{11}{1--25}.
\newblock
\begin{APACrefDOI} \doi{10.1029/2001GC000282} \end{APACrefDOI}
\PrintBackRefs{\CurrentBib}

\bibitem [\protect \citeauthoryear {%
Maclennan%
, Mckenzie%
\BCBL {}\ \BBA {} Gronv{\"o}ld%
}{%
Maclennan%
\ \protect \BOthers {.}}{%
{\protect \APACyear {2001}}%
}]{%
Maclennan2001}
\APACinsertmetastar {%
Maclennan2001}%
\begin{APACrefauthors}%
Maclennan, J.%
, Mckenzie, D.%
\BCBL {}\ \BBA {} Gronv{\"o}ld, K.%
\end{APACrefauthors}%
\unskip\
\newblock
\APACrefYearMonthDay{2001}{}{}.
\newblock
{\BBOQ}\APACrefatitle {{Plume-driven upwelling under central Iceland}}
  {{Plume-driven upwelling under central Iceland}}.{\BBCQ}
\newblock
\APACjournalVolNumPages{Earth and Planetary Science Letters}{194}{1-2}{67--82}.
\newblock
\begin{APACrefDOI} \doi{10.1016/S0012-821X(01)00553-2} \end{APACrefDOI}
\PrintBackRefs{\CurrentBib}

\bibitem [\protect \citeauthoryear {%
Maclennan%
\ \protect \BOthers {.}}{%
Maclennan%
\ \protect \BOthers {.}}{%
{\protect \APACyear {2003}}%
}]{%
Maclennan2003}
\APACinsertmetastar {%
Maclennan2003}%
\begin{APACrefauthors}%
Maclennan, J.%
, McKenzie, D.%
, Gronv{\"o}ld, K.%
, Shimizu, N.%
, Eiler, J\BPBI M.%
\BCBL {}\ \BBA {} Kitchen, N.%
\end{APACrefauthors}%
\unskip\
\newblock
\APACrefYearMonthDay{2003}{}{}.
\newblock
{\BBOQ}\APACrefatitle {{Melt mixing and crystallization under Theistareykir,
  northeast Iceland}} {{Melt mixing and crystallization under Theistareykir,
  northeast Iceland}}.{\BBCQ}
\newblock
\APACjournalVolNumPages{Geochemistry, Geophysics, Geosystems}{4}{11}{}.
\newblock
\begin{APACrefDOI} \doi{10.1029/2003GC000558} \end{APACrefDOI}
\PrintBackRefs{\CurrentBib}

\bibitem [\protect \citeauthoryear {%
Macpherson%
, Hilton%
, Day%
, Lowry%
\BCBL {}\ \BBA {} Gr{\"o}nvold%
}{%
Macpherson%
\ \protect \BOthers {.}}{%
{\protect \APACyear {2005}}%
}]{%
Macpherson2005}
\APACinsertmetastar {%
Macpherson2005}%
\begin{APACrefauthors}%
Macpherson, C\BPBI G.%
, Hilton, D\BPBI R.%
, Day, J\BPBI M.%
, Lowry, D.%
\BCBL {}\ \BBA {} Gr{\"o}nvold, K.%
\end{APACrefauthors}%
\unskip\
\newblock
\APACrefYearMonthDay{2005}{}{}.
\newblock
{\BBOQ}\APACrefatitle {{High-{$^{3}$}He/{$^{4}$}He, depleted mantle and
  low-$\delta${$^{18}$}O, recycled oceanic lithosphere in the source of central
  Iceland magmatism}} {{High-{$^{3}$}He/{$^{4}$}He, depleted mantle and
  low-$\delta${$^{18}$}O, recycled oceanic lithosphere in the source of central
  Iceland magmatism}}.{\BBCQ}
\newblock
\APACjournalVolNumPages{Earth and Planetary Science Letters}{233}{3}{411--427}.
\newblock
\begin{APACrefDOI} \doi{10.1016/j.epsl.2005.02.037} \end{APACrefDOI}
\PrintBackRefs{\CurrentBib}

\bibitem [\protect \citeauthoryear {%
McKenzie%
}{%
McKenzie%
}{%
{\protect \APACyear {1984}}%
}]{%
McKenzie1984}
\APACinsertmetastar {%
McKenzie1984}%
\begin{APACrefauthors}%
McKenzie, D.%
\end{APACrefauthors}%
\unskip\
\newblock
\APACrefYearMonthDay{1984}{}{}.
\newblock
{\BBOQ}\APACrefatitle {{The Generation and Compaction of Partially Molten
  Rock}} {{The Generation and Compaction of Partially Molten Rock}}.{\BBCQ}
\newblock
\APACjournalVolNumPages{Journal of Petrology}{25}{3}{713--765}.
\newblock
\begin{APACrefDOI} \doi{10.1093/petrology/25.3.713} \end{APACrefDOI}
\PrintBackRefs{\CurrentBib}

\bibitem [\protect \citeauthoryear {%
McKenzie%
\ \BBA {} Bickle%
}{%
McKenzie%
\ \BBA {} Bickle%
}{%
{\protect \APACyear {1988}}%
}]{%
McKenzie1988}
\APACinsertmetastar {%
McKenzie1988}%
\begin{APACrefauthors}%
McKenzie, D.%
\BCBT {}\ \BBA {} Bickle, M\BPBI J.%
\end{APACrefauthors}%
\unskip\
\newblock
\APACrefYearMonthDay{1988}{}{}.
\newblock
{\BBOQ}\APACrefatitle {{The Volume and Composition of Melt Generated by
  Extension of the Lithosphere}} {{The Volume and Composition of Melt Generated
  by Extension of the Lithosphere}}.{\BBCQ}
\newblock
\APACjournalVolNumPages{Journal of Petrology}{29}{3}{625--679}.
\newblock
\begin{APACrefDOI} \doi{10.1093/petrology/29.3.625} \end{APACrefDOI}
\PrintBackRefs{\CurrentBib}

\bibitem [\protect \citeauthoryear {%
{McKenzie}%
\ \BBA {} {O'Nions}%
}{%
{McKenzie}%
\ \BBA {} {O'Nions}%
}{%
{\protect \APACyear {1991}}%
}]{%
McKenzie1991}
\APACinsertmetastar {%
McKenzie1991}%
\begin{APACrefauthors}%
{McKenzie}, D.%
\BCBT {}\ \BBA {} {O'Nions}, R\BPBI K.%
\end{APACrefauthors}%
\unskip\
\newblock
\APACrefYearMonthDay{1991}{{\APACmonth{10}}}{}.
\newblock
{\BBOQ}\APACrefatitle {{{Partial Melt Distributions from Inversion of Rare
  Earth Element Concentrations}}} {{{Partial Melt Distributions from Inversion
  of Rare Earth Element Concentrations}}}.{\BBCQ}
\newblock
\APACjournalVolNumPages{Journal of Petrology}{32}{}{1021--1091}.
\newblock
\begin{APACrefDOI} \doi{10.1093/petrology/32.5.1021} \end{APACrefDOI}
\PrintBackRefs{\CurrentBib}

\bibitem [\protect \citeauthoryear {%
Nicholson%
\ \protect \BOthers {.}}{%
Nicholson%
\ \protect \BOthers {.}}{%
{\protect \APACyear {1991}}%
}]{%
Nicholson1991}
\APACinsertmetastar {%
Nicholson1991}%
\begin{APACrefauthors}%
Nicholson, H.%
, Condomines, M.%
, Fitton, J\BPBI G.%
, Fallick, A\BPBI E.%
, Gr{\"{o}}nvold, K.%
\BCBL {}\ \BBA {} Rogers, G.%
\end{APACrefauthors}%
\unskip\
\newblock
\APACrefYearMonthDay{1991}{}{}.
\newblock
{\BBOQ}\APACrefatitle {{Geochemical and Isotopic Evidence for Crustal
  Assimilation Beneath Krafla, Iceland}} {{Geochemical and Isotopic Evidence
  for Crustal Assimilation Beneath Krafla, Iceland}}.{\BBCQ}
\newblock
\APACjournalVolNumPages{Journal of Petrology}{32}{5}{1005--1020}.
\newblock
\begin{APACrefDOI} \doi{10.1093/petrology/32.5.1005} \end{APACrefDOI}
\PrintBackRefs{\CurrentBib}

\bibitem [\protect \citeauthoryear {%
Nielsen%
\ \protect \BOthers {.}}{%
Nielsen%
\ \protect \BOthers {.}}{%
{\protect \APACyear {2007}}%
}]{%
Nielsen2007}
\APACinsertmetastar {%
Nielsen2007}%
\begin{APACrefauthors}%
Nielsen, S\BPBI G.%
, Rehk{\"{a}}mper, M.%
, Brandon, A\BPBI D.%
, Norman, M\BPBI D.%
, Turner, S.%
\BCBL {}\ \BBA {} O'Reilly, S\BPBI Y.%
\end{APACrefauthors}%
\unskip\
\newblock
\APACrefYearMonthDay{2007}{}{}.
\newblock
{\BBOQ}\APACrefatitle {{Thallium isotopes in Iceland and Azores lavas —
  Implications for the role of altered crust and mantle geochemistry}}
  {{Thallium isotopes in Iceland and Azores lavas — Implications for the role
  of altered crust and mantle geochemistry}}.{\BBCQ}
\newblock
\APACjournalVolNumPages{Earth and Planetary Science
  Letters}{264}{1-2}{332--345}.
\newblock
\begin{APACrefDOI} \doi{10.1016/j.epsl.2007.10.008} \end{APACrefDOI}
\PrintBackRefs{\CurrentBib}

\bibitem [\protect \citeauthoryear {%
Paterson%
}{%
Paterson%
}{%
{\protect \APACyear {1994}}%
}]{%
Paterson1994}
\APACinsertmetastar {%
Paterson1994}%
\begin{APACrefauthors}%
Paterson, W.%
\end{APACrefauthors}%
\unskip\
\newblock
\APACrefYear{1994}.
\newblock
\APACrefbtitle {{The Physics of Glaciers}} {{The Physics of Glaciers}}\
  (\PrintOrdinal{Third}\ \BEd).
\newblock
\APACaddressPublisher{Amsterdam}{Pergamon}.
\newblock
\begin{APACrefDOI} \doi{10.1016/B978-0-08-037944-9.50017-0} \end{APACrefDOI}
\PrintBackRefs{\CurrentBib}

\bibitem [\protect \citeauthoryear {%
Patton%
, Hubbard%
, Bradwell%
\BCBL {}\ \BBA {} Schomacker%
}{%
Patton%
\ \protect \BOthers {.}}{%
{\protect \APACyear {2017}}%
}]{%
Patton2017}
\APACinsertmetastar {%
Patton2017}%
\begin{APACrefauthors}%
Patton, H.%
, Hubbard, A.%
, Bradwell, T.%
\BCBL {}\ \BBA {} Schomacker, A.%
\end{APACrefauthors}%
\unskip\
\newblock
\APACrefYearMonthDay{2017}{}{}.
\newblock
{\BBOQ}\APACrefatitle {{The configuration, sensitivity and rapid retreat of the
  Late Weichselian Icelandic ice sheet}} {{The configuration, sensitivity and
  rapid retreat of the Late Weichselian Icelandic ice sheet}}.{\BBCQ}
\newblock
\APACjournalVolNumPages{Earth-Science Reviews}{166}{}{223--245}.
\newblock
\begin{APACrefDOI} \doi{10.1016/j.earscirev.2017.02.001} \end{APACrefDOI}
\PrintBackRefs{\CurrentBib}

\bibitem [\protect \citeauthoryear {%
Peate%
\ \protect \BOthers {.}}{%
Peate%
\ \protect \BOthers {.}}{%
{\protect \APACyear {2009}}%
}]{%
Peate2009}
\APACinsertmetastar {%
Peate2009}%
\begin{APACrefauthors}%
Peate, D\BPBI W.%
, Baker, J\BPBI A.%
, Jakobsson, S\BPBI P.%
, Waight, T\BPBI E.%
, Kent, A\BPBI J.%
, Grassineau, N\BPBI V.%
\BCBL {}\ \BBA {} Skovgaard, A\BPBI C.%
\end{APACrefauthors}%
\unskip\
\newblock
\APACrefYearMonthDay{2009}{}{}.
\newblock
{\BBOQ}\APACrefatitle {{Historic magmatism on the Reykjanes Peninsula, Iceland:
  A snap-shot of melt generation at a ridge segment}} {{Historic magmatism on
  the Reykjanes Peninsula, Iceland: A snap-shot of melt generation at a ridge
  segment}}.{\BBCQ}
\newblock
\APACjournalVolNumPages{Contributions to Mineralogy and
  Petrology}{157}{3}{359--382}.
\newblock
\begin{APACrefDOI} \doi{10.1007/s00410-008-0339-4} \end{APACrefDOI}
\PrintBackRefs{\CurrentBib}

\bibitem [\protect \citeauthoryear {%
Peate%
\ \protect \BOthers {.}}{%
Peate%
\ \protect \BOthers {.}}{%
{\protect \APACyear {2010}}%
}]{%
Peate2010}
\APACinsertmetastar {%
Peate2010}%
\begin{APACrefauthors}%
Peate, D\BPBI W.%
, Breddam, K.%
, Baker, J\BPBI A.%
, Kurz, M\BPBI D.%
, Barker, A\BPBI K.%
, Prestvik, T.%
\BDBL {}Skovgaard, A\BPBI C.%
\end{APACrefauthors}%
\unskip\
\newblock
\APACrefYearMonthDay{2010}{}{}.
\newblock
{\BBOQ}\APACrefatitle {{Compositional characteristics and spatial distribution
  of enriched Icelandic mantle components}} {{Compositional characteristics and
  spatial distribution of enriched Icelandic mantle components}}.{\BBCQ}
\newblock
\APACjournalVolNumPages{Journal of Petrology}{51}{7}{1447--1475}.
\newblock
\begin{APACrefDOI} \doi{10.1093/petrology/egq025} \end{APACrefDOI}
\PrintBackRefs{\CurrentBib}

\bibitem [\protect \citeauthoryear {%
P{\'e}tursson%
, Nor{\dh}dahl%
\BCBL {}\ \BBA {} Ing{\'o}lfsson%
}{%
P{\'e}tursson%
\ \protect \BOthers {.}}{%
{\protect \APACyear {2015}}%
}]{%
Petursson2015}
\APACinsertmetastar {%
Petursson2015}%
\begin{APACrefauthors}%
P{\'e}tursson, H.%
, Nor{\dh}dahl, H.%
\BCBL {}\ \BBA {} Ing{\'o}lfsson, {\'O}.%
\end{APACrefauthors}%
\unskip\
\newblock
\APACrefYearMonthDay{2015}{}{}.
\newblock
{\BBOQ}\APACrefatitle {{Late Weichselian history of relative sea level changes
  in Iceland during a collapse and subsequent retreat of marine based ice
  sheet}} {{Late Weichselian history of relative sea level changes in Iceland
  during a collapse and subsequent retreat of marine based ice sheet}}.{\BBCQ}
\newblock
\APACjournalVolNumPages{Cuadernos de Investigación
  Geogr{\'a}fica}{41}{2}{261--277}.
\newblock
\begin{APACrefDOI} \doi{10.18172/cig.2741} \end{APACrefDOI}
\PrintBackRefs{\CurrentBib}

\bibitem [\protect \citeauthoryear {%
Poreda%
, Schilling%
\BCBL {}\ \BBA {} Craig%
}{%
Poreda%
\ \protect \BOthers {.}}{%
{\protect \APACyear {1986}}%
}]{%
Poreda1986}
\APACinsertmetastar {%
Poreda1986}%
\begin{APACrefauthors}%
Poreda, R.%
, Schilling, J\BHBI G.%
\BCBL {}\ \BBA {} Craig, H.%
\end{APACrefauthors}%
\unskip\
\newblock
\APACrefYearMonthDay{1986}{}{}.
\newblock
{\BBOQ}\APACrefatitle {{Helium and hydrogen isotopes in ocean-ridge basalts
  north and south of Iceland}} {{Helium and hydrogen isotopes in ocean-ridge
  basalts north and south of Iceland}}.{\BBCQ}
\newblock
\APACjournalVolNumPages{Earth and Planetary Science Letters}{78}{1}{1--17}.
\newblock
\begin{APACrefDOI} \doi{10.1016/0012-821X(86)90168-8} \end{APACrefDOI}
\PrintBackRefs{\CurrentBib}

\bibitem [\protect \citeauthoryear {%
Rudge%
, Maclennan%
\BCBL {}\ \BBA {} Stracke%
}{%
Rudge%
\ \protect \BOthers {.}}{%
{\protect \APACyear {2013}}%
}]{%
Rudge2013}
\APACinsertmetastar {%
Rudge2013}%
\begin{APACrefauthors}%
Rudge, J\BPBI F.%
, Maclennan, J.%
\BCBL {}\ \BBA {} Stracke, A.%
\end{APACrefauthors}%
\unskip\
\newblock
\APACrefYearMonthDay{2013}{}{}.
\newblock
{\BBOQ}\APACrefatitle {{The geochemical consequences of mixing melts from a
  heterogeneous mantle}} {{The geochemical consequences of mixing melts from a
  heterogeneous mantle}}.{\BBCQ}
\newblock
\APACjournalVolNumPages{Geochimica et Cosmochimica Acta}{114}{}{112--143}.
\newblock
\begin{APACrefDOI} \doi{10.1016/j.gca.2013.03.042} \end{APACrefDOI}
\PrintBackRefs{\CurrentBib}

\bibitem [\protect \citeauthoryear {%
S{\ae}mundsson%
}{%
S{\ae}mundsson%
}{%
{\protect \APACyear {1991}}%
}]{%
Saemundsson1991}
\APACinsertmetastar {%
Saemundsson1991}%
\begin{APACrefauthors}%
S{\ae}mundsson, K.%
\end{APACrefauthors}%
\unskip\
\newblock
\APACrefYearMonthDay{1991}{}{}.
\newblock
{\BBOQ}\APACrefatitle {{Jardfr{\ae}di Kr{\"o}flukerfisins. In: Gardarsson A,
  Einarsson A (eds) N{\'a}tt{\'u}ra M{\'y}vatns}} {{Jardfr{\ae}di
  Kr{\"o}flukerfisins. In: Gardarsson A, Einarsson A (eds) N{\'a}tt{\'u}ra
  M{\'y}vatns}}.{\BBCQ}
\newblock
\APACjournalVolNumPages{Hid islenska n{\'a}tt{\'u}rufr{\ae}dif{\'e}lg,
  Reykjav{\'i}k}{}{}{24--95}.
\PrintBackRefs{\CurrentBib}

\bibitem [\protect \citeauthoryear {%
S{\ae}mundsson%
\ \protect \BOthers {.}}{%
S{\ae}mundsson%
\ \protect \BOthers {.}}{%
{\protect \APACyear {2012}}%
}]{%
Saemundsson2012}
\APACinsertmetastar {%
Saemundsson2012}%
\begin{APACrefauthors}%
S{\ae}mundsson, K.%
, Hjartarson, {\'A}.%
, Kaldal, I.%
, Sigurgeirsson, M\BPBI {\'A}.%
, Kristinsson, S\BPBI G.%
\BCBL {}\ \BBA {} V{\'i}kingsson, S.%
\end{APACrefauthors}%
\unskip\
\newblock
\APACrefYearMonthDay{2012}{}{}.
\newblock
{\BBOQ}\APACrefatitle {{Geological Map of Northern Volcanic Zone, Iceland.
  Northern Part. 1:100 000. Reykjav{\'i}k: Iceland GeoSurvey and
  Landsvirkjun.}} {{Geological Map of Northern Volcanic Zone, Iceland. Northern
  Part. 1:100 000. Reykjav{\'i}k: Iceland GeoSurvey and Landsvirkjun.}}{\BBCQ}
\newblock

\PrintBackRefs{\CurrentBib}

\bibitem [\protect \citeauthoryear {%
S{\ae}mundsson%
, Sigurgeirsson%
, Hjartarson%
, Kaldal%
\BCBL {}\ \BBA {} Kristinsson%
}{%
S{\ae}mundsson%
\ \protect \BOthers {.}}{%
{\protect \APACyear {2016}}%
}]{%
Saemundsson2016}
\APACinsertmetastar {%
Saemundsson2016}%
\begin{APACrefauthors}%
S{\ae}mundsson, K.%
, Sigurgeirsson, M\BPBI {\'A}.%
, Hjartarson, {\'A}.%
, Kaldal, I.%
\BCBL {}\ \BBA {} Kristinsson, S\BPBI G.%
\end{APACrefauthors}%
\unskip\
\newblock
\APACrefYearMonthDay{2016}{}{}.
\newblock
{\BBOQ}\APACrefatitle {{Geological Map of of Southwest Iceland, 1:100 000 (2nd
  ed.). Reykjav{\'i}k: Iceland GeoSurvey.}} {{Geological Map of of Southwest
  Iceland, 1:100 000 (2nd ed.). Reykjav{\'i}k: Iceland GeoSurvey.}}{\BBCQ}
\newblock

\PrintBackRefs{\CurrentBib}

\bibitem [\protect \citeauthoryear {%
Shaw%
}{%
Shaw%
}{%
{\protect \APACyear {1970}}%
}]{%
Shaw1970}
\APACinsertmetastar {%
Shaw1970}%
\begin{APACrefauthors}%
Shaw, D\BPBI M.%
\end{APACrefauthors}%
\unskip\
\newblock
\APACrefYearMonthDay{1970}{}{}.
\newblock
{\BBOQ}\APACrefatitle {{Trace element fractionation during anatexis}} {{Trace
  element fractionation during anatexis}}.{\BBCQ}
\newblock
\APACjournalVolNumPages{Geochimica et Cosmochimica Acta}{34}{2}{237--243}.
\newblock
\begin{APACrefDOI} \doi{10.1016/0016-7037(70)90009-8} \end{APACrefDOI}
\PrintBackRefs{\CurrentBib}

\bibitem [\protect \citeauthoryear {%
Sigmundsson%
}{%
Sigmundsson%
}{%
{\protect \APACyear {1991}}%
}]{%
Sigmundsson1991}
\APACinsertmetastar {%
Sigmundsson1991}%
\begin{APACrefauthors}%
Sigmundsson, F.%
\end{APACrefauthors}%
\unskip\
\newblock
\APACrefYearMonthDay{1991}{}{}.
\newblock
{\BBOQ}\APACrefatitle {{Post-glacial rebound and asthenosphere viscosity in
  Iceland}} {{Post-glacial rebound and asthenosphere viscosity in
  Iceland}}.{\BBCQ}
\newblock
\APACjournalVolNumPages{Geophysical Research Letters}{18}{6}{1131--1134}.
\newblock
\begin{APACrefDOI} \doi{10.1029/91GL01342} \end{APACrefDOI}
\PrintBackRefs{\CurrentBib}

\bibitem [\protect \citeauthoryear {%
Sigurdsson%
, Schilling%
\BCBL {}\ \BBA {} Meyer%
}{%
Sigurdsson%
\ \protect \BOthers {.}}{%
{\protect \APACyear {1978}}%
}]{%
Sigurdsson1978}
\APACinsertmetastar {%
Sigurdsson1978}%
\begin{APACrefauthors}%
Sigurdsson, H.%
, Schilling, J\BHBI G.%
\BCBL {}\ \BBA {} Meyer, P\BPBI S.%
\end{APACrefauthors}%
\unskip\
\newblock
\APACrefYearMonthDay{1978}{}{}.
\newblock
{\BBOQ}\APACrefatitle {{Skagi and Langj{\"{o}}kull Volcanic Zones in Iceland:
  1. Petrology and structure}} {{Skagi and Langj{\"{o}}kull Volcanic Zones in
  Iceland: 1. Petrology and structure}}.{\BBCQ}
\newblock
\APACjournalVolNumPages{Journal of Geophysical Research}{83}{B8}{3971}.
\newblock
\begin{APACrefDOI} \doi{10.1029/JB083iB08p03971} \end{APACrefDOI}
\PrintBackRefs{\CurrentBib}

\bibitem [\protect \citeauthoryear {%
Sigvaldason%
, Annertz%
\BCBL {}\ \BBA {} Nilsson%
}{%
Sigvaldason%
\ \protect \BOthers {.}}{%
{\protect \APACyear {1992}}%
}]{%
Sigvaldason1992}
\APACinsertmetastar {%
Sigvaldason1992}%
\begin{APACrefauthors}%
Sigvaldason, G\BPBI E.%
, Annertz, K.%
\BCBL {}\ \BBA {} Nilsson, M.%
\end{APACrefauthors}%
\unskip\
\newblock
\APACrefYearMonthDay{1992}{}{}.
\newblock
{\BBOQ}\APACrefatitle {{Effect of glacier loading/deloading on volcanism:
  postglacial volcanic production rate of the Dyngjufj{\"{o}}ll area, central
  Iceland}} {{Effect of glacier loading/deloading on volcanism: postglacial
  volcanic production rate of the Dyngjufj{\"{o}}ll area, central
  Iceland}}.{\BBCQ}
\newblock
\APACjournalVolNumPages{Bulletin of Volcanology}{54}{5}{385--392}.
\newblock
\begin{APACrefDOI} \doi{10.1007/BF00312320} \end{APACrefDOI}
\PrintBackRefs{\CurrentBib}

\bibitem [\protect \citeauthoryear {%
Sims%
\ \protect \BOthers {.}}{%
Sims%
\ \protect \BOthers {.}}{%
{\protect \APACyear {2013}}%
}]{%
Sims2013}
\APACinsertmetastar {%
Sims2013}%
\begin{APACrefauthors}%
Sims, K\BPBI W.%
, Maclennan, J.%
, Blichert-Toft, J.%
, Mervine, E\BPBI M.%
, Blusztajn, J.%
\BCBL {}\ \BBA {} Grönvold, K.%
\end{APACrefauthors}%
\unskip\
\newblock
\APACrefYearMonthDay{2013}{}{}.
\newblock
{\BBOQ}\APACrefatitle {{Short length scale mantle heterogeneity beneath Iceland
  probed by glacial modulation of melting}} {{Short length scale mantle
  heterogeneity beneath Iceland probed by glacial modulation of
  melting}}.{\BBCQ}
\newblock
\APACjournalVolNumPages{Earth and Planetary Science Letters}{379}{}{146--157}.
\newblock
\begin{APACrefDOI} \doi{10.1016/j.epsl.2013.07.027} \end{APACrefDOI}
\PrintBackRefs{\CurrentBib}

\bibitem [\protect \citeauthoryear {%
Sinton%
, Gr{\"{o}}nvold%
\BCBL {}\ \BBA {} S{\ae}mundsson%
}{%
Sinton%
\ \protect \BOthers {.}}{%
{\protect \APACyear {2005}}%
}]{%
Sinton2005}
\APACinsertmetastar {%
Sinton2005}%
\begin{APACrefauthors}%
Sinton, J.%
, Gr{\"{o}}nvold, K.%
\BCBL {}\ \BBA {} S{\ae}mundsson, K.%
\end{APACrefauthors}%
\unskip\
\newblock
\APACrefYearMonthDay{2005}{}{}.
\newblock
{\BBOQ}\APACrefatitle {{Postglacial eruptive history of the Western Volcanic
  Zone, Iceland}} {{Postglacial eruptive history of the Western Volcanic Zone,
  Iceland}}.{\BBCQ}
\newblock
\APACjournalVolNumPages{Geochemistry, Geophysics, Geosystems}{6}{12}{}.
\newblock
\begin{APACrefDOI} \doi{10.1029/2005GC001021} \end{APACrefDOI}
\PrintBackRefs{\CurrentBib}

\bibitem [\protect \citeauthoryear {%
Skovgaard%
, Storey%
, Baker%
, Blusztajn%
\BCBL {}\ \BBA {} Hart%
}{%
Skovgaard%
\ \protect \BOthers {.}}{%
{\protect \APACyear {2001}}%
}]{%
Skovgaard2001}
\APACinsertmetastar {%
Skovgaard2001}%
\begin{APACrefauthors}%
Skovgaard, A\BPBI C.%
, Storey, M.%
, Baker, J.%
, Blusztajn, J.%
\BCBL {}\ \BBA {} Hart, S\BPBI R.%
\end{APACrefauthors}%
\unskip\
\newblock
\APACrefYearMonthDay{2001}{}{}.
\newblock
{\BBOQ}\APACrefatitle {{Osmium-oxygen isotopic evidence for a recycled and
  strongly depleted component in the Iceland mantle plume}} {{Osmium-oxygen
  isotopic evidence for a recycled and strongly depleted component in the
  Iceland mantle plume}}.{\BBCQ}
\newblock
\APACjournalVolNumPages{Earth and Planetary Science
  Letters}{194}{1-2}{259--275}.
\newblock
\begin{APACrefDOI} \doi{10.1016/S0012-821X(01)00549-0} \end{APACrefDOI}
\PrintBackRefs{\CurrentBib}

\bibitem [\protect \citeauthoryear {%
Slater%
, Jull%
, McKenzie%
\BCBL {}\ \BBA {} Gronv{\"{o}}ld%
}{%
Slater%
\ \protect \BOthers {.}}{%
{\protect \APACyear {1998}}%
}]{%
Slater1998}
\APACinsertmetastar {%
Slater1998}%
\begin{APACrefauthors}%
Slater, L.%
, Jull, M.%
, McKenzie, D.%
\BCBL {}\ \BBA {} Gronv{\"{o}}ld, K.%
\end{APACrefauthors}%
\unskip\
\newblock
\APACrefYearMonthDay{1998}{}{}.
\newblock
{\BBOQ}\APACrefatitle {{Deglaciation effects on mantle melting under Iceland:
  Results from the northern volcanic zone}} {{Deglaciation effects on mantle
  melting under Iceland: Results from the northern volcanic zone}}.{\BBCQ}
\newblock
\APACjournalVolNumPages{Earth and Planetary Science
  Letters}{164}{1-2}{151--164}.
\newblock
\begin{APACrefDOI} \doi{10.1016/S0012-821X(98)00200-3} \end{APACrefDOI}
\PrintBackRefs{\CurrentBib}

\bibitem [\protect \citeauthoryear {%
Slater%
, McKenzie%
, Gr{\"o}nvold%
\BCBL {}\ \BBA {} Shimizu%
}{%
Slater%
\ \protect \BOthers {.}}{%
{\protect \APACyear {2001}}%
}]{%
Slater2001}
\APACinsertmetastar {%
Slater2001}%
\begin{APACrefauthors}%
Slater, L.%
, McKenzie, D.%
, Gr{\"o}nvold, K.%
\BCBL {}\ \BBA {} Shimizu, N.%
\end{APACrefauthors}%
\unskip\
\newblock
\APACrefYearMonthDay{2001}{}{}.
\newblock
{\BBOQ}\APACrefatitle {{Melt Generation and Movement beneath Theistareykir, NE
  Iceland}} {{Melt Generation and Movement beneath Theistareykir, NE
  Iceland}}.{\BBCQ}
\newblock
\APACjournalVolNumPages{Journal of Petrology}{42}{2}{321--354}.
\newblock
\begin{APACrefDOI} \doi{10.1093/petrology/42.2.321} \end{APACrefDOI}
\PrintBackRefs{\CurrentBib}

\bibitem [\protect \citeauthoryear {%
Smith%
\ \BBA {} Asimow%
}{%
Smith%
\ \BBA {} Asimow%
}{%
{\protect \APACyear {2005}}%
}]{%
Smith2005}
\APACinsertmetastar {%
Smith2005}%
\begin{APACrefauthors}%
Smith, P\BPBI M.%
\BCBT {}\ \BBA {} Asimow, P\BPBI D.%
\end{APACrefauthors}%
\unskip\
\newblock
\APACrefYearMonthDay{2005}{}{}.
\newblock
{\BBOQ}\APACrefatitle {{Adiabat\_1ph: A new public front-end to the MELTS,
  pMELTS, and pHMELTS models}} {{Adiabat\_1ph: A new public front-end to the
  MELTS, pMELTS, and pHMELTS models}}.{\BBCQ}
\newblock
\APACjournalVolNumPages{Geochemistry, Geophysics, Geosystems}{6}{2}{}.
\newblock
\begin{APACrefDOI} \doi{10.1029/2004GC000816} \end{APACrefDOI}
\PrintBackRefs{\CurrentBib}

\bibitem [\protect \citeauthoryear {%
Sobolev%
, Hofmann%
, Br{\"u}gmann%
, Batanova%
\BCBL {}\ \BBA {} Kuzmin%
}{%
Sobolev%
\ \protect \BOthers {.}}{%
{\protect \APACyear {2008}}%
}]{%
Sobolev2008}
\APACinsertmetastar {%
Sobolev2008}%
\begin{APACrefauthors}%
Sobolev, A\BPBI V.%
, Hofmann, A\BPBI W.%
, Br{\"u}gmann, G.%
, Batanova, V\BPBI G.%
\BCBL {}\ \BBA {} Kuzmin, D\BPBI V.%
\end{APACrefauthors}%
\unskip\
\newblock
\APACrefYearMonthDay{2008}{}{}.
\newblock
{\BBOQ}\APACrefatitle {{A Quantitative Link Between Recycling and Osmium
  Isotopes}} {{A Quantitative Link Between Recycling and Osmium
  Isotopes}}.{\BBCQ}
\newblock
\APACjournalVolNumPages{Science}{321}{5888}{536--536}.
\newblock
\begin{APACrefDOI} \doi{10.1126/science.1158452} \end{APACrefDOI}
\PrintBackRefs{\CurrentBib}

\bibitem [\protect \citeauthoryear {%
S{\'o}lnes%
, {\'A}sgeirsson%
, Bessason%
\BCBL {}\ \BBA {} Sigmundsson%
}{%
S{\'o}lnes%
\ \protect \BOthers {.}}{%
{\protect \APACyear {2013}}%
}]{%
solnes2013}
\APACinsertmetastar {%
solnes2013}%
\begin{APACrefauthors}%
S{\'o}lnes, J.%
, {\'A}sgeirsson, {\'A}.%
, Bessason, B.%
\BCBL {}\ \BBA {} Sigmundsson, F.%
\end{APACrefauthors}%
\unskip\
\newblock
\APACrefYearMonthDay{2013}{}{}.
\newblock
{\BBOQ}\APACrefatitle {{Reykjav{\'i}k: Vi{\dh}lagatrygging/
  H{\'a}sk{\'o}la{\'u}tg{\'a}fan}} {{Reykjav{\'i}k: Vi{\dh}lagatrygging/
  H{\'a}sk{\'o}la{\'u}tg{\'a}fan}}.{\BBCQ}
\newblock
\APACjournalVolNumPages{N{\'a}tt{\'u}ruv{\'a} {\'A} {\'I}slandi, Eldgos og
  Jar{\dh}skj{\'a}lftar}{}{}{}.
\PrintBackRefs{\CurrentBib}

\bibitem [\protect \citeauthoryear {%
Spiegelman%
}{%
Spiegelman%
}{%
{\protect \APACyear {1996}}%
}]{%
Spiegelman1996}
\APACinsertmetastar {%
Spiegelman1996}%
\begin{APACrefauthors}%
Spiegelman, M.%
\end{APACrefauthors}%
\unskip\
\newblock
\APACrefYearMonthDay{1996}{}{}.
\newblock
{\BBOQ}\APACrefatitle {{Geochemical consequences of melt transport in 2-D: The
  sensitivity of trace elements to mantle dynamics}} {{Geochemical consequences
  of melt transport in 2-D: The sensitivity of trace elements to mantle
  dynamics}}.{\BBCQ}
\newblock
\APACjournalVolNumPages{Earth and Planetary Science Letters}{139}{1}{115--132}.
\newblock
\begin{APACrefDOI} \doi{10.1016/0012-821X(96)00008-8} \end{APACrefDOI}
\PrintBackRefs{\CurrentBib}

\bibitem [\protect \citeauthoryear {%
Spiegelman%
\ \BBA {} McKenzie%
}{%
Spiegelman%
\ \BBA {} McKenzie%
}{%
{\protect \APACyear {1987}}%
}]{%
Spiegelman1987}
\APACinsertmetastar {%
Spiegelman1987}%
\begin{APACrefauthors}%
Spiegelman, M.%
\BCBT {}\ \BBA {} McKenzie, D.%
\end{APACrefauthors}%
\unskip\
\newblock
\APACrefYearMonthDay{1987}{}{}.
\newblock
{\BBOQ}\APACrefatitle {{Simple 2-D models for melt extraction at mid-ocean
  ridges and island arcs}} {{Simple 2-D models for melt extraction at mid-ocean
  ridges and island arcs}}.{\BBCQ}
\newblock
\APACjournalVolNumPages{Earth and Planetary Science
  Letters}{83}{1-4}{137--152}.
\newblock
\begin{APACrefDOI} \doi{10.1016/0012-821X(87)90057-4} \end{APACrefDOI}
\PrintBackRefs{\CurrentBib}

\bibitem [\protect \citeauthoryear {%
Stracke%
\ \protect \BOthers {.}}{%
Stracke%
\ \protect \BOthers {.}}{%
{\protect \APACyear {2003}}%
}]{%
Stracke2003}
\APACinsertmetastar {%
Stracke2003}%
\begin{APACrefauthors}%
Stracke, A.%
, Zindler, A.%
, Salters, V\BPBI J.%
, McKenzie, D.%
, Janne, B\BPBI T.%
, Albar{\`{e}}de, F.%
\BCBL {}\ \BBA {} Gr{\"{o}}nvold, K.%
\end{APACrefauthors}%
\unskip\
\newblock
\APACrefYearMonthDay{2003}{}{}.
\newblock
{\BBOQ}\APACrefatitle {{Theistareykir revisited}} {{Theistareykir
  revisited}}.{\BBCQ}
\newblock
\APACjournalVolNumPages{Geochemistry, Geophysics, Geosystems}{4}{2}{}.
\newblock
\begin{APACrefDOI} \doi{10.1029/2001GC000201} \end{APACrefDOI}
\PrintBackRefs{\CurrentBib}

\bibitem [\protect \citeauthoryear {%
Swindles%
\ \protect \BOthers {.}}{%
Swindles%
\ \protect \BOthers {.}}{%
{\protect \APACyear {2017}}%
}]{%
Swindles2017}
\APACinsertmetastar {%
Swindles2017}%
\begin{APACrefauthors}%
Swindles, G\BPBI T.%
, Watson, E\BPBI J.%
, Savov, I\BPBI P.%
, Lawson, I\BPBI T.%
, Schmidt, A.%
, Hooper, A.%
\BDBL {}Carrivick, J\BPBI L.%
\end{APACrefauthors}%
\unskip\
\newblock
\APACrefYearMonthDay{2017}{}{}.
\newblock
{\BBOQ}\APACrefatitle {{Climatic control on Icelandic volcanic activity during
  the mid-Holocene}} {{Climatic control on Icelandic volcanic activity during
  the mid-Holocene}}.{\BBCQ}
\newblock
\APACjournalVolNumPages{Geology}{46}{1}{47--50}.
\newblock
\begin{APACrefDOI} \doi{10.1130/G39633.1} \end{APACrefDOI}
\PrintBackRefs{\CurrentBib}

\bibitem [\protect \citeauthoryear {%
Thirlwall%
\ \protect \BOthers {.}}{%
Thirlwall%
\ \protect \BOthers {.}}{%
{\protect \APACyear {2006}}%
}]{%
Thirlwall2006}
\APACinsertmetastar {%
Thirlwall2006}%
\begin{APACrefauthors}%
Thirlwall, M\BPBI F.%
, Gee, M\BPBI A.%
, Lowry, D.%
, Mattey, D\BPBI P.%
, Murton, B\BPBI J.%
\BCBL {}\ \BBA {} Taylor, R\BPBI N.%
\end{APACrefauthors}%
\unskip\
\newblock
\APACrefYearMonthDay{2006}{}{}.
\newblock
{\BBOQ}\APACrefatitle {{Low $\delta${$^{18}$}O in the Icelandic mantle and its
  origins: Evidence from Reykjanes Ridge and Icelandic lavas}} {{Low
  $\delta${$^{18}$}O in the Icelandic mantle and its origins: Evidence from
  Reykjanes Ridge and Icelandic lavas}}.{\BBCQ}
\newblock
\APACjournalVolNumPages{Geochimica et Cosmochimica Acta}{70}{4}{993--1019}.
\newblock
\begin{APACrefDOI} \doi{10.1016/j.gca.2005.09.008} \end{APACrefDOI}
\PrintBackRefs{\CurrentBib}

\bibitem [\protect \citeauthoryear {%
Thirlwall%
, Gee%
, Taylor%
\BCBL {}\ \BBA {} Murton%
}{%
Thirlwall%
\ \protect \BOthers {.}}{%
{\protect \APACyear {2004}}%
}]{%
Thirlwall2004}
\APACinsertmetastar {%
Thirlwall2004}%
\begin{APACrefauthors}%
Thirlwall, M\BPBI F.%
, Gee, M\BPBI A.%
, Taylor, R\BPBI N.%
\BCBL {}\ \BBA {} Murton, B\BPBI J.%
\end{APACrefauthors}%
\unskip\
\newblock
\APACrefYearMonthDay{2004}{}{}.
\newblock
{\BBOQ}\APACrefatitle {{Mantle components in Iceland and adjacent ridges
  investigated using double-spike Pb isotope ratios}} {{Mantle components in
  Iceland and adjacent ridges investigated using double-spike Pb isotope
  ratios}}.{\BBCQ}
\newblock
\APACjournalVolNumPages{Geochimica et Cosmochimica Acta}{68}{2}{361--386}.
\newblock
\begin{APACrefDOI} \doi{10.1016/S0016-7037(03)00424-1} \end{APACrefDOI}
\PrintBackRefs{\CurrentBib}

\bibitem [\protect \citeauthoryear {%
White%
, McKenzie%
\BCBL {}\ \BBA {} O'Nions%
}{%
White%
\ \protect \BOthers {.}}{%
{\protect \APACyear {1992}}%
}]{%
White1992}
\APACinsertmetastar {%
White1992}%
\begin{APACrefauthors}%
White, R\BPBI S.%
, McKenzie, D.%
\BCBL {}\ \BBA {} O'Nions, R\BPBI K.%
\end{APACrefauthors}%
\unskip\
\newblock
\APACrefYearMonthDay{1992}{}{}.
\newblock
{\BBOQ}\APACrefatitle {{Oceanic crustal thickness from seismic measurements and
  rare earth element inversions}} {{Oceanic crustal thickness from seismic
  measurements and rare earth element inversions}}.{\BBCQ}
\newblock
\APACjournalVolNumPages{Journal of Geophysical Research: Solid
  Earth}{97}{B13}{19683--19715}.
\newblock
\begin{APACrefDOI} \doi{10.1029/92JB01749} \end{APACrefDOI}
\PrintBackRefs{\CurrentBib}

\bibitem [\protect \citeauthoryear {%
Workman%
\ \BBA {} Hart%
}{%
Workman%
\ \BBA {} Hart%
}{%
{\protect \APACyear {2005}}%
}]{%
Workman2005}
\APACinsertmetastar {%
Workman2005}%
\begin{APACrefauthors}%
Workman, R\BPBI K.%
\BCBT {}\ \BBA {} Hart, S\BPBI R.%
\end{APACrefauthors}%
\unskip\
\newblock
\APACrefYearMonthDay{2005}{}{}.
\newblock
{\BBOQ}\APACrefatitle {{Major and trace element composition of the depleted
  MORB mantle (DMM)}} {{Major and trace element composition of the depleted
  MORB mantle (DMM)}}.{\BBCQ}
\newblock
\APACjournalVolNumPages{Earth and Planetary Science Letters}{231}{1-2}{53--72}.
\newblock
\begin{APACrefDOI} \doi{10.1016/j.epsl.2004.12.005} \end{APACrefDOI}
\PrintBackRefs{\CurrentBib}

\end{thebibliography}
%




\end{document}